\pdfoutput=1
\documentclass{article}
\usepackage{etex}
\reserveinserts{28}

\usepackage[utf8]{inputenc}

\usepackage[pdftex]{graphicx}

\usepackage[in]{fullpage}

\usepackage{multicol}
\usepackage{multirow}
\usepackage{amsmath}
\usepackage{amssymb}
\usepackage{float}
\usepackage{textcomp}
\usepackage{tikz}
\usepackage{IEEEtrantools}
\usepackage{pgfplots}
\usepackage{verbatim}
\usepackage{schemabloc}
\usepackage{latexsym}
\usepackage{lscape}
\usepackage{subfigure}
\usepackage{yfonts}
\usepackage{mathabx}

\usepackage{longtable}
\usepackage{tabularx}
\usepackage{listings}
\usepackage{array}
\usepackage[refpages]{gloss}
\usepackage{textgreek}
\usepackage[numbers, sort, square]{natbib} 

\definecolor{LinkColor}{rgb}{0,0,0}
\usepackage{hyperref}

\graphicspath{{./pics/}}

\newcommand{\autor}{Dr. Jos\'e Alonso \textbf{Sol\'is-Lemus}$^1$, 
        Tiffany Baptiste$^1$, Rosie Barrows$^1$, Charles Sillett$^1$, \\ 
        Dr. Ali Gharaviri$^{1,3}$, 
        Giulia Raffaele$^1$, 
        Dr. Orod Razeghi$^{4,1}$, Dr. Marina Strocchi$^1$,\\
        Dr. Iain Sim$^1$, Dr. Irum Kotadia$^1$, Dr. Neil Bodagh$^1$, 
        Dr. Daniel O'Hare$^1$, \\ 
        Prof. Mark O'Neill$^1$, Dr. Steven E Williams$^{1,3}$, 
        Dr. Caroline Roney$^{1,2,6}$
        Prof. Steven Niederer$^{1,5,6}$}
\newcommand{\titulo}{Evaluation of an Open-Source Pipeline to Create Patient-Specific 
                    Left Atrial Models: A Reproducibility Study}

\author{\autor}
\title{\titulo}
\date{\begin{flushleft}
    \small
    1. School of Biomedical Engineering and Imaging Sciences, King's College London\\
    2. Queen Mary University of London \\
    3. Centre for Cardiovascular Science, University of Edinburgh\\ 
    4. Department of Haematology, NHS Blood and Transplant Centre, University of Cambridge\\
    5. Alan Turing Institute\\ 
    6.~\emph{Both collaborators serving as joint last-authors}
\end{flushleft}}

\begin{document}
\maketitle
\begin{abstract}
  This work presents an open-source software pipeline to create patient-specific
  left atrial models with fibre orientations and a fibrosis map,
  suitable for electrophysiology simulations, and
  quantifies the intra and inter observer reproducibility of the model creation. 
  The semi-automatic pipeline takes as input a contrast enhanced magnetic resonance 
  angiogram, and a late gadolinium enhanced (LGE) contrast magnetic resonance (CMR).
  Five operators were allocated 20 cases each from a set of 50 CMR datasets
  to create a total of 100 models to evaluate inter and intra-operator variability.
  Each output model consisted of:
  (1) a labelled surface mesh open at the pulmonary veins and mitral valve,
  (2) fibre orientations mapped from a diffusion tensor MRI (DTMRI) human atlas,
  (3) fibrosis map extracted from the LGE-CMR scan,
  and (4) simulation of local activation time (LAT) and phase singularity (PS) mapping.
  Reproducibility in our pipeline was evaluated by comparing agreement in shape of the output meshes,
  fibrosis distribution in the left atrial body, and fibre orientations.
  Reproducibility in simulations outputs was evaluated in the LAT maps by 
  comparing the total activation times, and the mean conduction velocity (CV). 
  PS maps were compared with the structural similarity index measure (SSIM).
  The users processed in total 60 cases for inter and
  40 cases for intra-operator variability.
  Our workflow allows a single model to be created in 16.72 $\pm$ 12.25 minutes.
  Similarity was measured with shape, percentage of fibres oriented in the same direction,
  and intra-class correlation coefficient (ICC) for the fibrosis calculation.
  Shape differed noticeably only with users' selection of the mitral valve and
  the length of the pulmonary veins from the ostia to the distal end;
  fibrosis agreement was high, with ICC of 0.909 (inter) and 0.999 (intra);
  fibre orientation agreement was high with 60.63\% (inter) and 71.77\% (intra).
  The LAT showed good agreement,
  where the median $\pm$ IQR of the absolute difference of the total activation times
  was 2.02 $\pm$ 2.45 ms for inter, and 1.37 $\pm$ 2.45 ms for intra.
  Also, the average $\pm$ sd of the mean CV  difference was
  -0.00404 $\pm$ 0.0155 $m/s$ for inter, and 0.0021 $\pm$ 0.0115 $m/s$ for intra.
  Finally, the PS maps showed a moderately good agreement in SSIM for inter and intra, where
  the mean $\pm$ sd SSIM for inter and intra were
  0.648 $\pm$ 0.21 and 0.608 $\pm$ 0.15, respectively.
  Although we found notable differences in the models,
  as a consequence of user input, our tests show that the uncertainty caused by
  both inter and intra-operator variability is comparable with uncertainty
  due to estimated fibres, and image resolution accuracy of segmentation tools.
\end{abstract}
\newpage
\paragraph{Research highlights:}
\begin{itemize}
  \item New open-source pipeline creates patient-specific left atrial models from CMR scans.
  \item Inter/intra-operator variability evaluated in study with 100 models.
  \item Model building time reduced to just 16.7 minutes.
  \item Error measurements from operator variability comparable to differences observed with varied Image resolution or fibre field.
  \end{itemize}
  

\newpage
\section{Introduction}
Patient-specific computational models of the heart are moving from a research tool
to industrial and clinical applications~\cite{pras2018-intro-regulatory}.
Regulatory bodies and societies are now providing verification, validation and uncertainty quantification
frameworks for the evaluation of these models, and the steps and tests required to demonstrate
model credibility for their context of use~\cite{pras2021-intro-credibility}.

In the area of patient-specific cardiac models, previous research has addressed areas of
code verification providing N-version and analytical benchmark
problems~\cite{Niederer2011_electrophysiology,pras2014_verification,sander2015},
availability of independent validation data sets, 
and adoption of uncertainty quantification techniques~\cite{Pathmanathan2018}.
However, little attention has been given to the uncertainty introduced by operator 
decisions in patient-specific cardiac modelling workflows. 
In particular, the impact of \textbf{intra-operator variability}, which refers to 
the variation in measurements made by the same operator when performing a task 
multiple times, and \textbf{inter-operator variability}, which refers to the 
variation in measurements made by different operators when performing the same 
task, has not been adequately addressed.
This knowledge gap is particularly relevant considering the routine reporting of 
intra and inter-observer variability in medical imaging~\cite{moon2002-intro-reprod,alabed2022-intro-mri}. 

The high cost in model creation to date and manual workflows have limited the evaluation of 
model creation reproducibility.
Performing inter and intra-observer variability studies is bound by the 
capacity to analyse sufficiently large data sets by different operators.
However, the capacity to create such data sets is currently limited, as past modeling 
studies have built only a few patient-specific models, often fewer than 10. 
While more recent studies have included 20 to 50 patients~\cite{Lee2019_gender}, 
only a few studies with 80 or more cases have been 
reported~\cite{Roney2022-predicting,bifulco2021_90models}.
To enable and motivate studies on intra and inter-observer variability, 
it is necessary to develop robust software platforms that facilitate the creation of 
larger cohorts of models.

CemrgApp~\cite{Razeghi2020-cemrgapp} is an open-source platform,
aggregating different open source workflows,
for performing medical image analysis and creating patient-specific models.
Specifically, CemrgApp provides a tool for left atrial anatomical
and structural cardiac MRI analysis, which measures an anatomy and estimates tissue fibrosis burden.
Multiple studies have been performed by us and others, which confirm the excellent
inter and intra-observer variability of the atrial image analysis workflow.
We have used this workflow in reproducibility assessment of atrial 
fibrosis~\cite{Sim2019-reproducibility},
evaluation of left atrial scar formation~\cite{oneill2019-pulmonary},
and optimisation of LGE-CMR imaging of post-ablation atrial scar~\cite{chubb2018-reproducibility}.
Independent image analysis studies have used the tool to verify that CemrgApp's atrial 
fibrosis analysis replicates the results from a third-party 
software~\cite{hopman2021-quantification,hopman2022_ra}. 
Furthermore, CemrgApp atrial scar analysis has been demonstrated in scans from different 
vendors (Siemens and Phillips) and validated with public image data bases~\cite{Razeghi2020-cnn}.
The output of the scar quantification tool for left atrial image analysis is
a surface mesh that can be analysed, imported into electro-anatomical mapping system,
or used as the basis for creating a patient-specific model.
The mesh with the estimated fibrosis burden can then be augmented with
a universal coordinate system and atrial fibres~\cite{Roney2019-uac},
derived from an atlas of ex-vivo human diffusion tensor MRI (DTMRI)~\cite{Roney2021-atlas}.
The fibrosis burden can be used to estimate tissue conduction and
cellular electrophysiology properties to generate a patient-specific model.
We have previously used this approach to generate cohorts of patient-specific
models~\cite{Roney2022-predicting}.
However, the current available tools for creating these models were often in different
software platforms, taking on average 4.5 hours of processing time 
per patient case~\cite{Roney2020-cinc}.

Despite the extensive research done in the area of patient-specific cardiac models,
a notable gap in the literature remains regarding reproducibility studies that address 
intra and inter-observer variability. 
This gap is important because accounting for uncertainty in model predictions due to operator 
variability is crucial for ensuring confidence in simulation predictions in clinical and 
regulatory applications.  
Furthermore, bridging the gap would impact existing modelling approaches' accuracy whether 
these models guide procedures~\citep{Boyle2021_af,Boyle2019_optima},
or their outputs are used as inputs to classifiers for predicting clinical 
outcomes~\cite{Roney2022-predicting}.
In this context, we \textbf{aim} to fill this gap by  
(i) introducing an open-source workflow for creating patient-specific
atrial models from cardiac MRI and 
(ii) performing an intra and inter-observer study to demonstrate the reproducibility 
in the approach for creating patient-specific left atrial models and 
quantify uncertainty in model predictions introduced by manual steps in model creation.
The \textbf{objective} of this study is to quantify the impact of intra and inter-observer
variability on atrial fibre maps, activation simulations, and fibrillation simulations.
\newpage
We demonstrate that with a guided, semi-automated modelling approach we can generate
operator-independent patient-specific left atrial models.
By providing the first evaluation of model reproducibility, 
we can provide estimates of the degree of uncertainty due to manual operations that give 
context for interpreting clinical and research simulation studies.
Section~\ref{sec:materials} describes the data and refers to the image acquisition
protocols, as well as the users, assignment of cases, and training resources developed. 
Section~\ref{sec:methods} describes the methodology, simulation protocols, and the 
reproducibility experiments and metrics evaluated. 
Section~\ref{sec:results} presents the results of all the experiments.

\begin{figure}[hbpt]
    \centering
    \includegraphics[width=\textwidth]{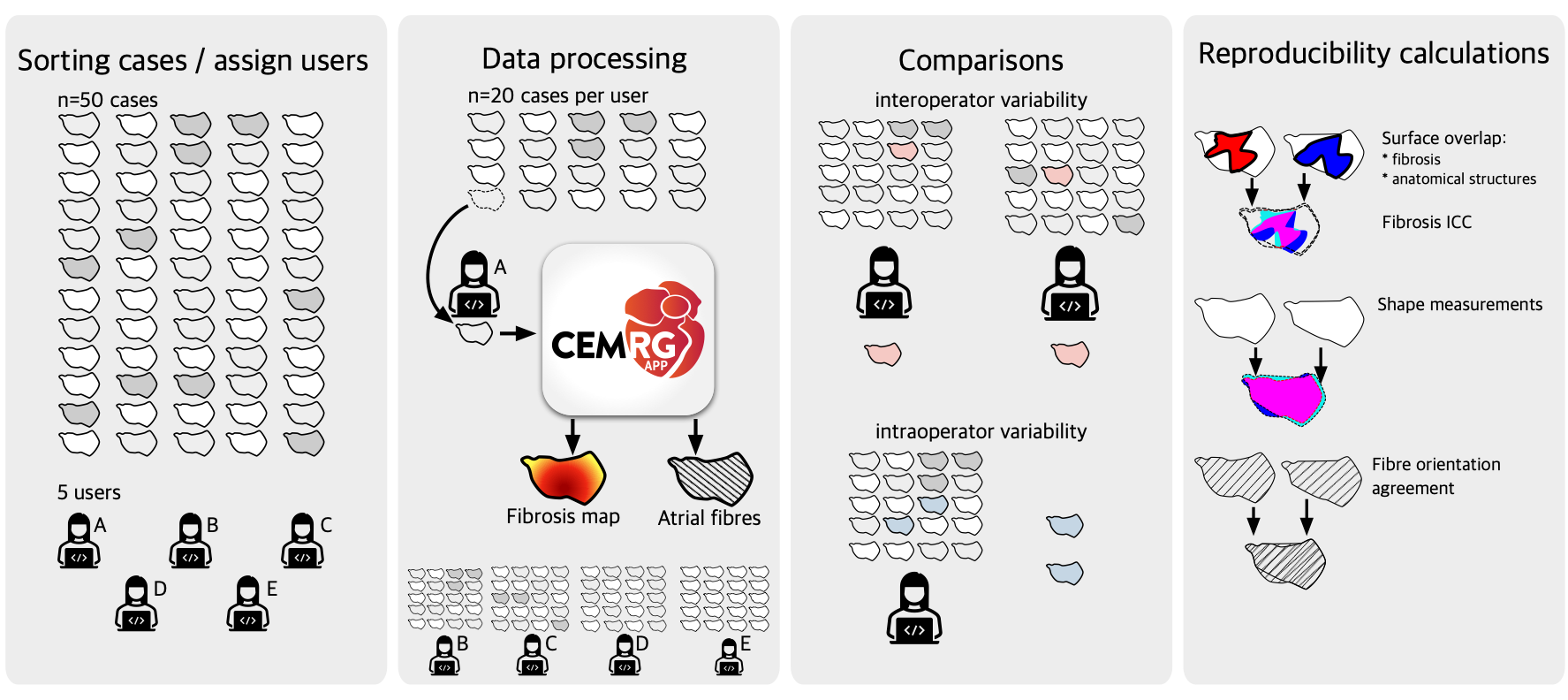}
    \caption[Overview of this study.]
    {Overview of this study.
    Five independent users processed 20 cases at random from a pool of 50 cases.
    Processing was done using the CemrgApp pipeline developed for this work,
    which involves the creation of a simulation-ready mesh, fibres orientations,
    and a fibrosis map.
    Users processed some cases twice to test for intra-operator variability,
    whilst other cases were processed by two users to test for inter-operator
    variability.
    The pipeline was assessed for reproducibility at various stages: the
    surface area overlap, shortest distance of each points, and fibre orientations.
    Each of the 100 output cases ($20\times5$ users) were used to run 3 electrophysiology
    simulations, where total activation time, absolute LAT differences, and
    correlation of PS map in universal atrial coordinates (UAC) were calculated.
    }\label{fig:intro-abstract}
\end{figure}
\newpage
\section{Materials}\label{sec:materials}
\begin{figure}[hbpt]
    \centering
    \includegraphics[width=\textwidth]{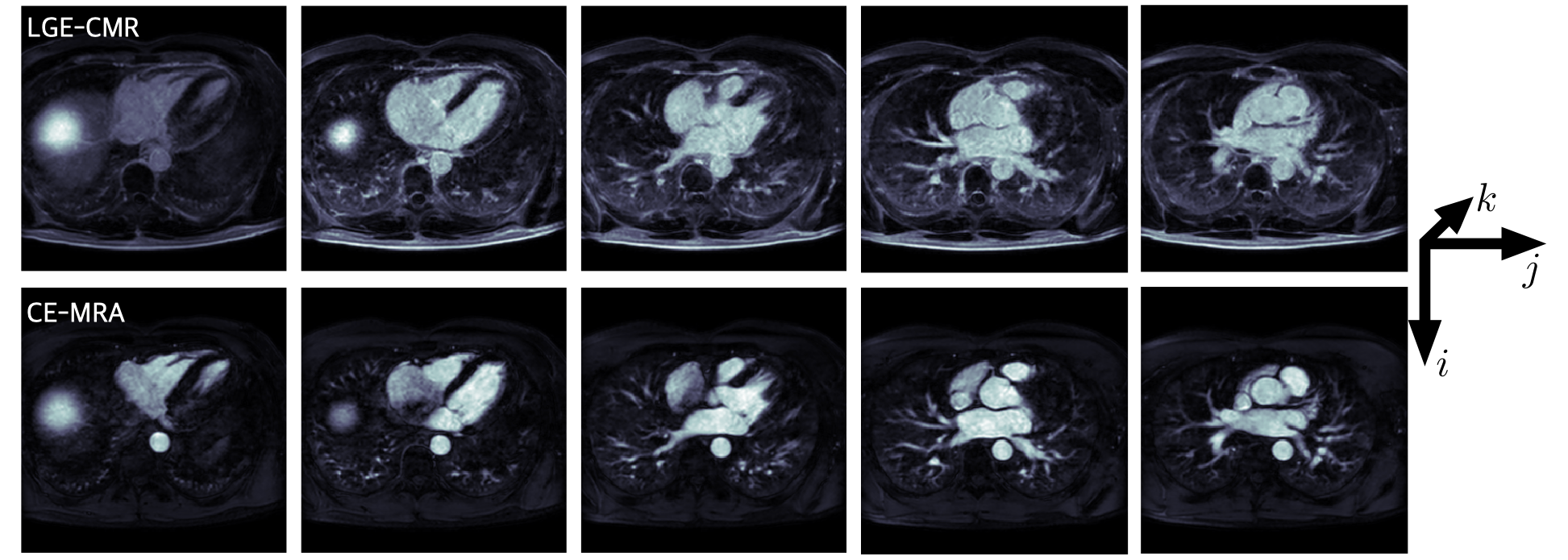}
    \caption[Example axial slices of input volumes.]
    {Example axial slices of input volumes.
    Each dataset consisted of an LGE-CMR scan (top) and its corresponding CE-MRA scan (bottom).
    Scans were screened to ensure the orientation is RAI
    (Right-to-left, Anterior-to-posterior, Inferior-to-superior).
    Scans were resampled to be isotropic, that is, each voxel had a resolution of
    $1\text{mm}^3$.}\label{fig:mat-data}
\end{figure}

Fifty cases were analysed, each consisting of two scans:
an ECG-triggered, contrast enhanced magnetic resonance angiogram (CE-MRA),
and late gadolinium enhanced cardiac magnetic resonance (LGE-CMR).
CMR imaging was performed on Phillips and Siemens 1.5T scanners~\cite{Sim2019-reproducibility}.
The full description of image acquisition is reported by~\cite{chubb2018-reproducibility}.
All input DICOM files were resampled to be isotropic, reoriented and stored in
the Nifti-2 format, which ensures data anonymisation.
Images were resampled so each voxel had a resolution of $1\text{mm}^3$.
All images were screened to fit in the \emph{RAI} orientation, which orients the
voxels in the X axis from Right-to-left, the Y axis from Anterior-to-posterior,
and the Z axis from Inferior-to-superior (Figure~\ref{fig:mat-data}).

\paragraph{Data Allocation and Training}
Data were divided amongst five users: the developer (d), three novice users (abc),
and one senior user with more experience with medical images (s).
Only the developer had seen and screened each of the cases 
before they were randomly assigned to each users.
Each user (abcd) was randomly assigned 20 cases to process.
The data assigned to each user were categorised as follows:
(i) \textbf{10 cases} for \textbf{intra}-operator variability;
(ii) \textbf{5 cases} for \textbf{inter}-operator variability vs. another user (abcd);
(iii) \textbf{5 cases} for inter-operator variability vs. senior user (s).
To quantify inter-operator variability, each of the thirty cases was assigned
twice.
For the intra-operator variability cases, each user had 5 cases independently
repeated from their own set.
The developer obtained full knowledge of which cases were repeated in each user's pool
only after model creation and analyses were performed.
Users received training resources such as videos of each stage of the data processing.
A standard operating procedure (SOP) document was written describing the whole pipeline.
Training and documentation are available as Supplementary Material.

\section{Methods}\label{sec:methods}
This study consisted of three stages of processing: an image-to-mesh analysis pipeline, 
fibre mapping through universal atrial coordinates, and electrophysiology simulations. 
The software pipeline and the reproducibility assessment of inter and intra-operator variability
along each of the three stages are the main contributions of this work.

The image-to-mesh analysis pipeline and fibre mapping were integrated 
into a single software workflow, which was 
developed in the CemrgApp framework (\cite{Razeghi2020-cemrgapp}),
an open-source platform to develop image analysis and computer vision workflows.
CemrgApp constitutes a platform used to develop standalone pipelines with a specific task.
Standalone pipelines are developed as a sequenced set of buttons called plugins.

For this study, a plugin was developed to streamline the creation of simulation-ready
meshes from a CMR scan.
The plugin developed involves three processing stages:
(1) conversion from a pair of scans to a labelled mesh,
(2) calculation of universal atrial coordinates through a docker container
(hosted at \url{hub.docker.com/repository/docker/cemrg/uac}), and
(3) mapping DTMRI fibres from an atlas, as reported by~\cite{Roney2021-atlas}.
The outputs from the developed pipeline are:
the Universal Atrial Coordinates,
a labelled mesh with fibres, and a mesh with the fibrosis projection.
Each user submitted their cases, which were screened for quality control.
The next stage in data processing is to run simulations in openCARP,
an open-source simulation environment for cardiac electrophysiology~\cite{Plank2021-opencarp}.

\begin{figure}[t]
    \centering
    \includegraphics[width=\textwidth]{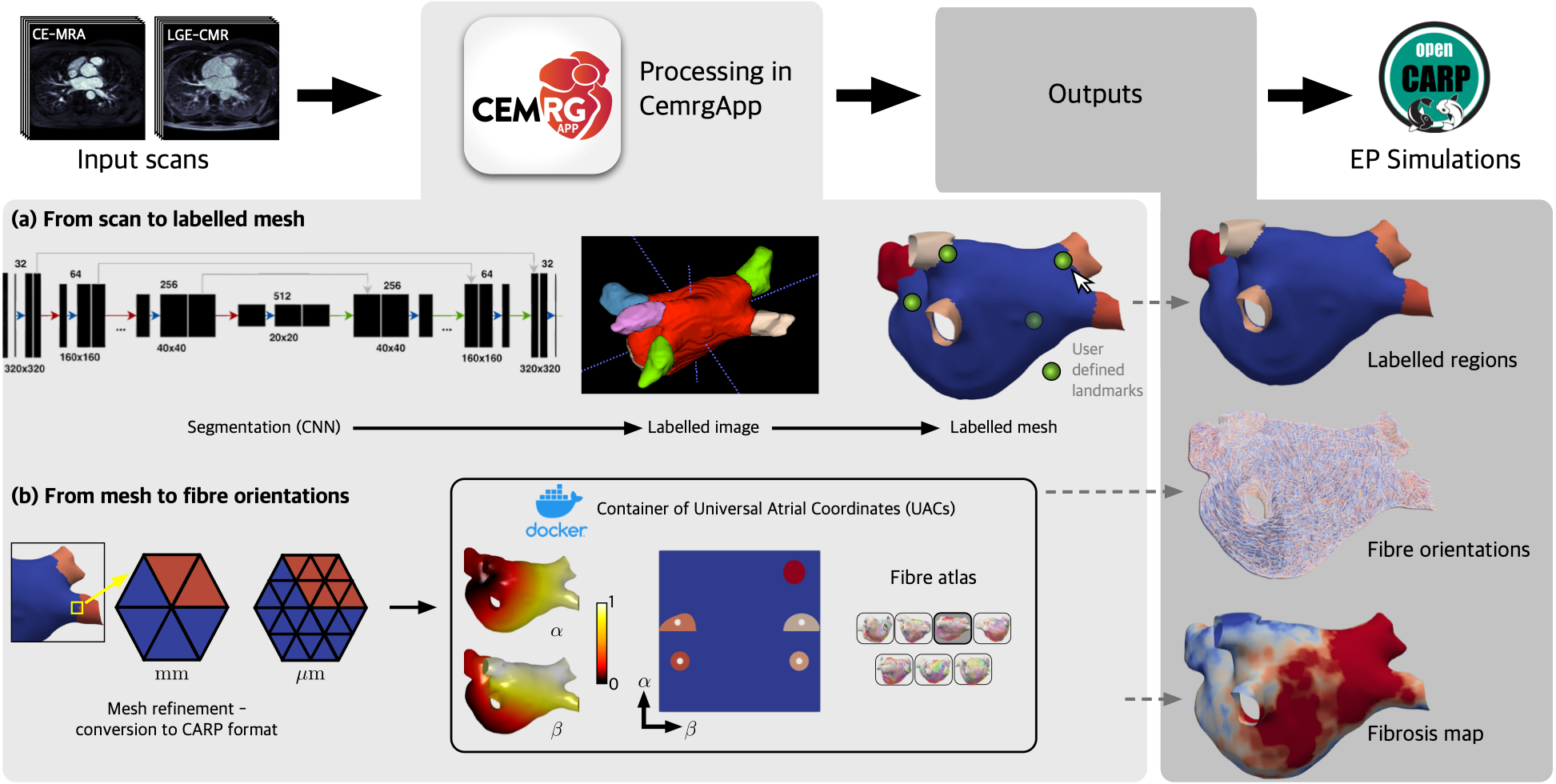}
    \caption[Overview of methodology to process a single MRA/LGE pair.]
    {Overview of methodology to process a single MRA/LGE pair.
    Scans are processed in CemrgApp through a combination of embedded and
    external code called through docker containers.
    The pipeline processes the scans (a) from segmentation to a labelled mesh,
    which is then (b) refined using meshtool (\cite{Neic2020-meshtool}) and processed
    with the Universal Atrial Coordinates (UAC) docker container.
    The UAC docker container creates a standardised frame of reference for the mesh,
    and projects DTMRI fibres from an atlas onto the mesh.
    Finally, the user produces a fibrosis map from the LGE signal intensity.
    Outputs produced per case are:
    a labelled mesh, files with fibre orientations, and a fibrosis map.
    Electrophysiological simulations are then run on openCARP.}\label{fig:methods-abstract}
\end{figure}

\subsection{Processing on CemrgApp}\label{sec:methods-cemrgapp}
This stage processes an input pair of scans, one optimised for atrial anatomy (CE-MRA),
one optimised for tissue characterisation (LGE-CMR), and outputs a labelled mesh.
A segmentation is produced from the input CE-MRA scan.
The segmentation is then registered to the LGE-CMR space,
to interrogate the fibrosis score of the left atrium.
Labelling of the segmentation is achieved by identifying
the pulmonary veins (PVs) and left atrial appendage (LAA).
From the labelled segmentation, a corresponding labelled surface mesh is produced.
Two modes of operation are available to the user in this stage of the pipeline:
\emph{semi-automatic} and \emph{manual}, described below.

\paragraph{Semi-automatic Pipeline.}
A multilabel segmentation of the atrium bloodpool is created using a convolutional
neural network (CNN) of the left atrium~\citet{Razeghi2020-cnn}.
The multilabel segmentation produces three distinct labels: the left atrial (LA) body,
pulmonary veins (PV) and left atrial appendage (LAA), and mitral valve (MV).
Naive labels are assigned to each of the pulmonary veins and
left atrial appendage to differentiate them.
Labels at this stage are assigned from the largest object to the smallest, 
thus these must later be identified by the user.
The surface mesh is generated and the naive labels projected onto it.

\paragraph{Manual Pipeline.}
The user segments the bloodpool of the CE-MRA manually in one of two ways:
using the single-label option of the CNN referenced before, or
using CemrgApp's fully manual segmentation module.
The single-label option of the CNN ignores the PV/LAA and MV labels created.
To standardise how the pulmonary veins and appendage are labelled, 
the user then identifies the pulmonary veins/appendage by setting
control points at the distal end of each vein/appendage.
Each point set selection prompts the user to identify the atrial structure they clicked:
left atrial body (LA), left atrial appendage (LAA), and pulmonary veins
left/right superior/inferior (LSPV, LIPV, RSPV, RIPV).
A line is drawn from each point to the centre of mass of the atrium,
where the radius of each PV or LAA is calculated and the inflection 
point of the radius is used to identify the transition of each 
atrial structure to the atrial body. 
The location is identified by a disk, which the user can move if necessary.
Once accepted, the veins/appendage are labelled according to the user's
selections using default label values.

\paragraph{Refining Mesh to be Simulation-ready.}
The following steps are the same for either mode of operation (whether manual or automatic).
Meshes at this stage should present six different labels: 
body (LA), appendage (LAA), and four pulmonary veins
(LSPV, LIPV, RSPV, RIPV). 
The user has the option to manually correct errors in the labelling of the surface mesh.
Furthermore, a label verification tool was developed to provide the user with
the option to automatically check and correct connectivity issues, 
for example, if some elements of the atrial body presented the label of 
the left atrial appendage.
The resulting mesh needs to be open at the pulmonary veins and mitral valve.
The user clips the mesh by choosing the centres and radii of spheres, 
which will clip the mesh at the distal ends of the pulmonary vein and 
at the mitral valve.
Once clipped, the mesh is refined to an average edge length of 0.3mm;
then it is cleaned from bad topology definitions, scaled to be in $\mu m$,
and converted to openCARP format using meshtool~\cite{Neic2020-meshtool}.

\paragraph{Fibrosis Map.}
The LGE-CMR scan is interrogated by projecting the maximum intensity of the
wall onto the surface elements of the clipped mesh.
The segmentation was done initially on the CE-MRA scan and registered to the LGE-CMR scan.
The wall intensities are estimated by superimposing the surface mesh on the scan
and calculate the maximum intensity projection of the voxels along the normal direction
of each element.
This creates a fibrosis map, to which a threshold can be applied, to find
areas with fibrosis or ablation scar.
A common technique to determine the threshold is the image-intensity ratio 
(IIR)~\cite{Khurram2014_iir}, 
in which the threshold is defined by the mean intensity of the bloodpool multiplied 
by a factor. 
Common threshold values are 
0.97~\cite{Sim2019-reproducibility}, 1.2, and 1.32~\cite{Benito2016_iir}.
The pipeline to obtain a fibrosis score has been extensively described before. 
The reader is referred to the works by 
\cite{Razeghi2020-cemrgapp,Razeghi2020-cnn,Sim2019-reproducibility,chubb2018-reproducibility}
for more detailed descriptions.

\subsection{Universal Atrial Coordinates in CermgApp.}\label{sec:methods-uac}
Universal atrial coordinates (UAC)~\cite{Roney2019-uac,Roney2021-atlas} constitute
a 2-dimensional frame of reference to compare different atrial geometries.
UAC are calculated by solving two Laplace equations with Dirichlet boundary conditions
defined by lines in the surface mesh of boundary nodes.
The coordinates, $(\alpha, \beta)$, are defined relative to the atrial structures:
pulmonary veins and left atrial appendage.
The first coordinate, $\alpha$, spans from the septal to the lateral walls;
the second coordinate, $\beta$, is defined from the posterior mitral valve,
over the roof to the anterior mitral valve.
Once the UAC process is finished, the selected atlas fibre files are mapped
onto the specific mesh.
The UAC software was packaged into a docker container~\cite{merkel2014docker},
and ran from within CemrgApp through its command line interface tool.
The container can be found through Docker's repository site at the
following link \url{https://hub.docker.com/repository/docker/cemrg/uac}.

\paragraph{User-selected Landmark Points.}
The user interface of the pipeline displays an interactive view of the mesh 
where the user selects the landmark points necessary for the calculation of the UAC.
Compared to previous implementations, the user is only required to
select four landmark points:
(1) at the junction between the left superior PV (LSPV) and the atrial body,
(2) at the junction between the right superior PV (RSPV) and the atrial body,
(3) on the lateral wall, between the LSPV, MV and LAA and
(4) on the septal wall, near the \emph{fossa ovalis} (FO).
The points are visible in Figure~\ref{fig:methods-abstract}.
Having less points selected reduces user input error in the UAC software.

\paragraph{Fibre Mapping.}
The user chooses which fibre orientations to project onto the mesh: 
epicardium, endocardium or both (bilayer).
The user can also select which fibre field from the atlas reported by~\cite{Roney2021-atlas}.
There are seven fibre files $(1,\cdots, 7)$, an average fibre field, $(a)$,
which aggregates the fibre orientations of all cases, 
and a rule-based fibre field $(l)$ by~\cite{labarthe2016_model}.
For this study, two fibre fields were mapped onto the processed meshes,
corresponding to DTMRI fibre file 1 and the Labarthe fibre file.

\subsection{Reproducibility Experiments from CemrgApp pipeline.}\label{sec:methods-reprod}
Reproducibility calculations were evaluated to assess variability of output from the CemrgApp pipeline.
Operator variability is assessed \textbf{between} (inter) users and \textbf{within} (intra) the same user.
Evaluations consisted of:
shape measurements, fibrosis agreement, and fibre orientation agreement.

\paragraph{Shape Measurements.}
The minimum euclidean distance
from any point in one mesh to the other was calculated.
Three measurements were made on the resulting array of minimal distances:
(i) Hausdorff distance~\cite{Acosta-Hausdorff}, defined as the
maximum of the array of minimal distances as a \emph{worst-case} scenario;
(ii) the mean of the minimal distance; and (iii) the median.

\paragraph{Fibrosis Agreement.}
This metric is assessed on the scar tissue defined by the surface where
the fibrosis signal, projected from the LGE-CMR, is above a pre-determined
threshold.
Fibrosis agreement is assessed through the intra-class correlation coefficient
(ICC)~\cite{Shrout-ICC,Elfving-ICC}, which assesses the reliability of ratings
by comparing the variability of different ratings of the same subject to
the total variation across all ratings and all subjects.
There are 6 different ways of calculating the ICC~\cite{Shrout-ICC}.
The variant of ICC calculated in this work is the average raters with absolute agreement.

\paragraph{Fibre Orientation Agreement.}
The comparison is done between the fibre orientations corresponding
to the closest elements between meshes.
The measurement to compare is the absolute value of the dot product between fibre orientations.
Note the direction of the fibres only needs to be co-lineal,
that is, angles between fibres of $0^{\circ}$ or $180^{\circ}$ are considered perfectly aligned.
The proportion of angle errors below $\pi/8 \quad (=22.5^{\circ})$ was calculated
as a measurement of fibre agreement~\cite{Roney2021-atlas}.

\subsection{Simulations}\label{sec:methods-simulations}
Two types of simulations were run on each of the 100 processed cases: 
baseline pacing to calculate local activation time (LAT) maps and 
atrial fibrillation simulations for which phase singularity (PS) maps
were calculated.
The openCARP simulator~\cite{Plank2021-opencarp} was used to run the simulations,
using the Courtemance human atrial model~\cite{courtemanche1998-ionic} with 
AF electrical remodelling~\cite{courtemanche1999ionic}.
Similar to the work by~\cite{Roney2021-atlas},
longitudinal conductivity was set to 0.4$S/m$ and
transverse conductivity to 0.1$S/m$. For baseline pacing, the model was stimulated at the RSPV rim and run for 1s. 
Local activation time (LAT) maps were calculated for bilayer model simulations
with two of the fibre field 1 and the Labarthe~\cite{labarthe2016_model} fibre field.
AF was initiated using an arrangement of four Archimedean spirals, 
and phase singularity maps were calculated following our previous study~\cite{roney2020silico}.

\paragraph{Assessment Metrics.}
The local activation time ($LAT$) maps were compared in two ways: 
first, the pairwise correlation coefficient was calculated from the mapping data, 
as reported by~\cite{Lubrecht2021_latcv}; 
then the total activation time was compared in both inter and intra-operator variability cases. 
The local conduction velocity (CV) was calculated as the inverse 
of the magnitude of the gradient ($1/\Vert \nabla LAT \Vert$)~\cite{Cantwell2015_cv}.
The mean conduction velocity was compared in both inter and intra-operator variability cases. 
Finally, as reported by~\cite{Li2020_ps}, 
two calculations were performed on each pair of phase singularity (PS) maps: 
the Pearson correlation coefficient and the structural similarity index. 


\section{Results}\label{sec:results}
Five users processed 100 cases in total,
where 40 correspond to intra-observer variability and 60 to
inter-observer variability.
After processing, users submitted their cases for assessment,
which were screened for quality control.
All meshes are available on Zenodo~\cite{solislemus2022_modrep_dataset}
at the link: \url{https://doi.org/10.5281/zenodo.7433015}.
Two cases presented a substantial user-error problem:
(i) the user identified the right pulmonary veins in incorrect order
during the mesh preprocessing stage of the \emph{semi-automatic pipeline};
(ii) another user forgot to clip the mesh's pulmonary veins before 
calculating the scar projection. More details can be found in the 
Supplementary Material.

\paragraph{Point/Element Correspondence for Comparisons.}
We created mappings between closest points and closest elements from meshes being compared
and the distance between these. We eliminated points farther than 1mm apart from each other.
Thus, every comparison apart from the shape agreement measurements is between points (or elements)
closer than 1mm.

\subsection{Reproducibility from the CemrgApp pipeline}
The mean and median times to complete the whole pipeline were 26 minutes and 25 seconds
and 16 minutes 43 seconds, respectively.
Most cases were completed between 14 and 36 minutes.
Four outliers were identified, where the process took 52, 100,
148, and 232 minutes respectively.
The hardware used varied from a laptop with 8GB RAM and 4 cores to a desktop workstation 
with  128GB RAM and 64 cores. 
In nine cases the time logging document was not created or deleted by the user.
Users attempted the \emph{semi-automatic} pipeline (Section~\ref{sec:methods-cemrgapp})
and only defaulted to the \emph{manual} pipeline when the multi-label segmentation
presented problems. 
Problems with the automatic segmentation were primarily related to 
the segmentation or labelling of the pulmonary veins.
For example, the neural network would output a good atrial body geometry, but 
presented missing or joined pulmonary veins which required manual intervention.
The distribution between semi-automatic and manual cases -out of 100- was 51 and 49, respectively.
None of the manual cases required the user to perform a fully manual segmentation.

\begin{figure}[hbpt]
    \includegraphics[width=\textwidth]{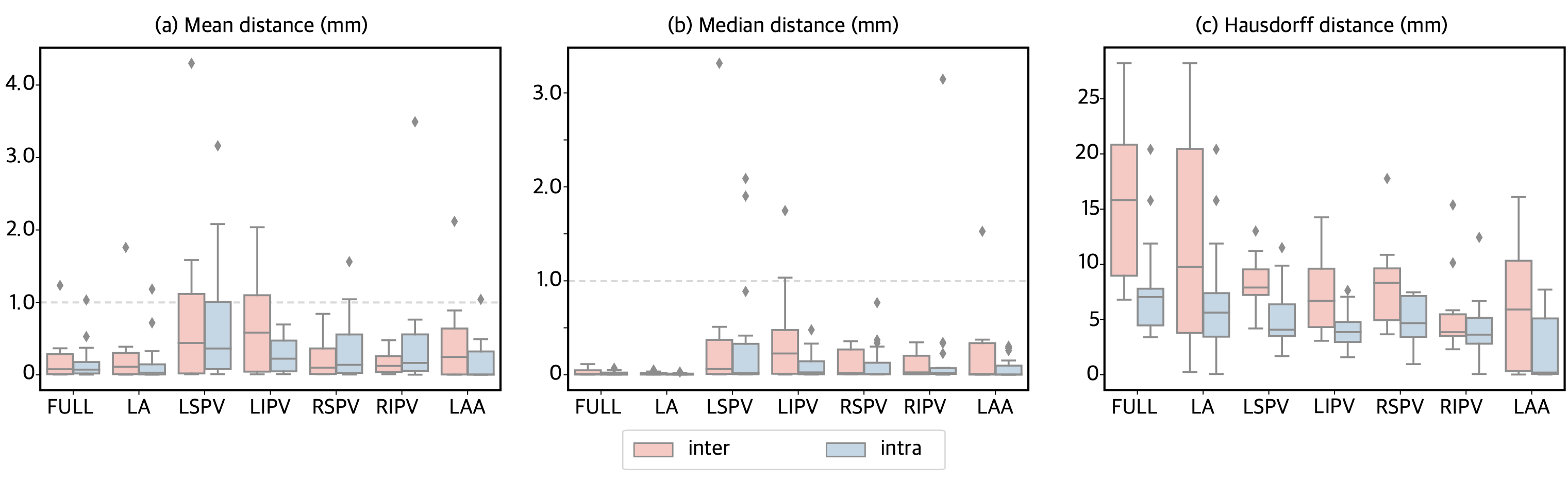}
    \caption[Distance to closest point metrics for inter and intra-operator]
    {Distance to closest point boxplots of the different metrics:
    (a) mean, (b) median, and (c) Hausdorff distance.
    Inter and intra-operator variability plots are shown in pink and blue, respectively.
    Different boxplots are presented to visualise the different structures:
    left atrial body (LA), left atrial appendage (LAA),
    as well as the pulmonary veins left superior (LSPV) and
    inferior (LIPV), and right superior (RSPV) and inferior
    Mean (a) and median (b) of the distance to closest point are overall under
    1mm.
    }\label{fig:res-inter-intra-dist}
\end{figure}

\paragraph{Shape Measurements.}
The calculations of Hausdorff distance, mean and median were calculated per atrial structure,
full atrium, left atrial body, left/right superior/inferior pulmonary veins,
and left atrial appendage.
Given the resolution of the image, an error of 1mm or less was considered \emph{acceptable}.
The majority of the mean and median smallest-distances were below 1 mm,
the left superior and inferior pulmonary veins presented the largest differences.
Only the left pulmonary veins reached values above 1mm in the mean distance,
although the 75th percentile were still calculated around or under 1mm.
The Hausdorff distance, as a \emph{worst case scenario} measurement,
had a mean value of 9.5 mm, but had ranges 
reaching almost 30mm in the inter-operator left atrial body
due to the uncertainty in the positioning of the mitral valve in the manual pipeline cases.
In all measurements, intra-operator variability results were notably smaller 
compared to the results corresponding to the inter-operator variability. 
Figure~\ref{fig:res-inter-intra-dist} shows boxplots showing the different measurements.

\begin{figure}[hbpt]
    \includegraphics[width=\textwidth]{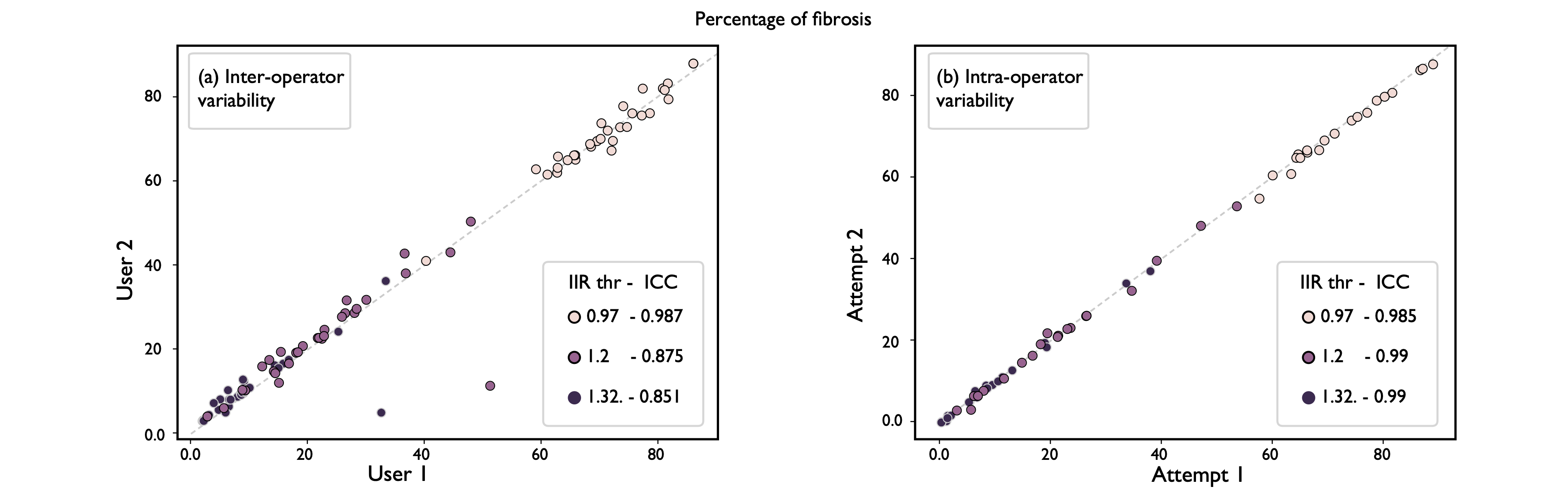}
    \caption[Fibrosis agreement]
    {Fibrosis agreement.
    Inter- (left) and intra-operator (right) variability are presented
    by showing the different fibrosis scores.
    On both axes represent the fibrosis score ranging from 0 to 1.
    Different colours represent different thresholds of the IIR method,
    which is presented next to the ICC coefficient.
    Points close to the identity line show good agreement.
    }\label{fig:res-fibrosis-agreement}
\end{figure}

\paragraph{Fibrosis Agreement.}
Figure~\ref{fig:res-fibrosis-agreement} shows the comparisons in fibrosis between
and within users.
We calculated ICC overall and per IIR threshold.
The IIR threshold is calculated by multiplying the mean of the bloodpool by 
the factor, in case of this study 0.97, 1.2, and 1.32. 
See Section~\ref{sec:methods-cemrgapp} for more details. 
On \textbf{inter-}operator variability comparisons ICC results were:
$ICC_{0.97}=0.987$, $ICC_{1.2}=0.875$, $ICC_{1.32}=0.851$ and $ICC_{all}=0.909$.
On \textbf{intra-}operator variability comparisons ICC results were:
$ICC_{0.97}=0.985$, $ICC_{1.2}=0.999$, $ICC_{1.32}=0.999$ and $ICC_{all}=0.999$.
We have included a table in the Supplementary Material, which
includes a full overview of the ICC coefficients,
p-values, and 95\% confidence intervals.


\paragraph{Fibre Orientation Agreement.}
The absolute value of the dot product,
a proxy for the angle between fibre orientations,
was calculated in pairs of meshes compared.
Elements in corresponding meshes at a greater distance than 1mm
were not added to the calculation,
given some of the differences in some meshes,
as seen in Figure~\ref{fig:res-inter-intra-dist}(c).
Results were separated into the different atrial structures and
stacked to showcase the distribution of values.
Values near 1 represent perfect alignment,
as it indicates an angle between fibre orientations of
$0^{\circ}$, or $180^{\circ}$.
To indicate \emph{good agreement}, a threshold of $22.5^{\circ}$ was chosen.
In the inter-operator variability pairings,
fibres in \emph{good agreement} represented approximately \textbf{60.63\%}
of all the fibres orientations across all cases.
In the intra-operator variability pairings,
the percentage of fibres in good agreement was approximately \textbf{71.77\%}.


\begin{figure}[hbpt]
  \centering
    \includegraphics[width=\textwidth]{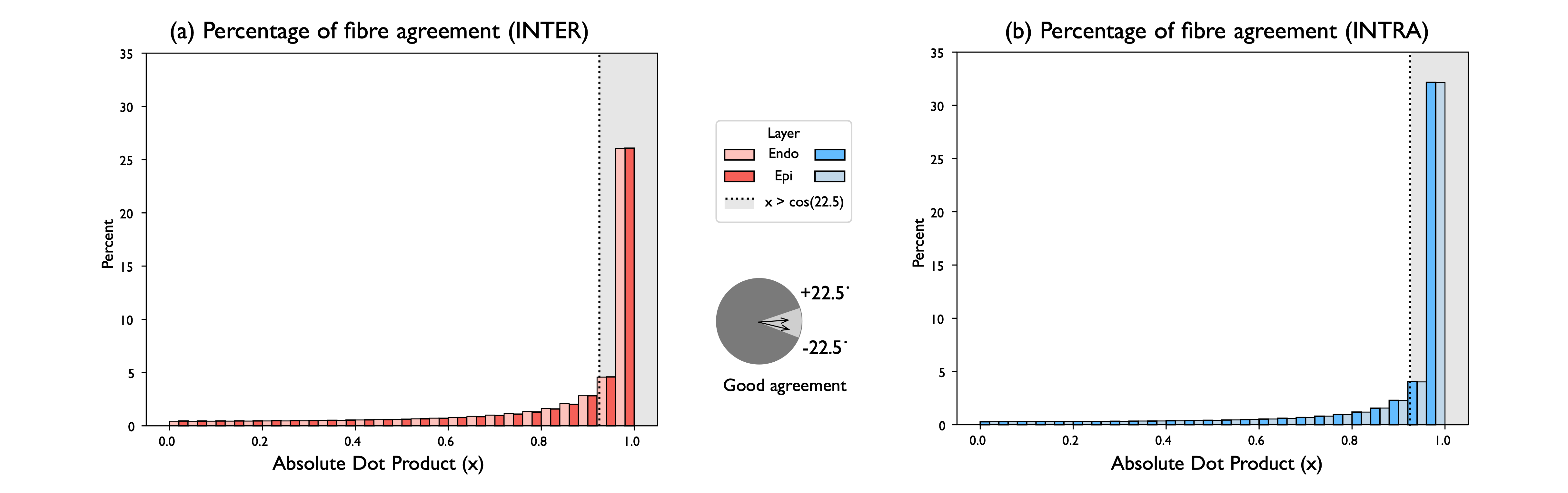}
    \caption[Fibre orientation agreement distribution.]
    {Fibre orientation agreement distribution.
      (a) Inter- and (b) intra-operator histograms of the distribution of the
      absolute value of the dot product between two fibres orientations.
      The histograms corresponding to the different atrial layers (endocardium and epicardium)
      have been distinguished to show relative distributions.
      Values greater than $\cos(22.5^{\circ})\approx 0.924$ correspond to angles
      between fibres between $\pm 22.5^{\circ}$,
      these were considered in \emph{good agreement}.
    }\label{fig:res-fibre-agreement}
\end{figure}

\subsection{Simulation Results}

\textbf{Local activation time (LAT) maps} had an excellent average correlation 
for both inter and intra comparisons, with mean ($\pm$ sd) correlations of 
0.992 ($\pm$ 0.007) and 0.996 ($\pm$ 0.003), respectively. 
Medians ($\pm$ IQR) of total activation times for inter and intra calculations were
131.31 ($\pm$ 18.59) and 132.61 ($\pm$ 19.08), respectively. 
The median ($\pm$ IQR) of the absolute difference of the total activation times was
2.02 ($\pm$ 2.45) ms for inter, and 1.37 ($\pm$ 2.45) ms for intra.
The Wilcoxon rank test for medians obtained a p-value of 0.86 and 0.83 for inter and intra, 
which suggests the test could not reject the hypothesis of equal medians in 
the distributions of total activation times. 

The \textbf{mean conduction velocity (CV)} was statistically different in almost all cases, 
where only 10 cases out of 50 (split as 5 in inter and 5 in intra) could not be determined 
as statistically different. 
In contrast, the average of the mean conduction velocity (CV) 
could not be determined as statistically different.
The average difference ($\pm$ sd) of the mean CV was 
-0.00404 ($\pm$ 0.0155) $m/s$ for inter, and 0.0021 ($\pm$ 0.0115) $m/s$ for intra comparisons.
The p-values for the comparisons between difference in average mean conduction velocity
 were found to be 0.535 for inter and 0.771 for intra. 
Finally, the \textbf{phase singularity maps} showed a mean ($\pm$ sd) correlation of 0.305 ($\pm$ 0.25) 
for the inter and 0.248 ($\pm$ 0.19) for the intra-operator variability comparisons. 
Regarding the structural similarity index, the mean ($\pm$ sd) values for inter and intra 
were higher at 0.648 ($\pm$ 0.21) and 0.608 ($\pm$ 0.15), respectively. 

\begin{figure}[t]
  \centering 
  \includegraphics[width=\textwidth]{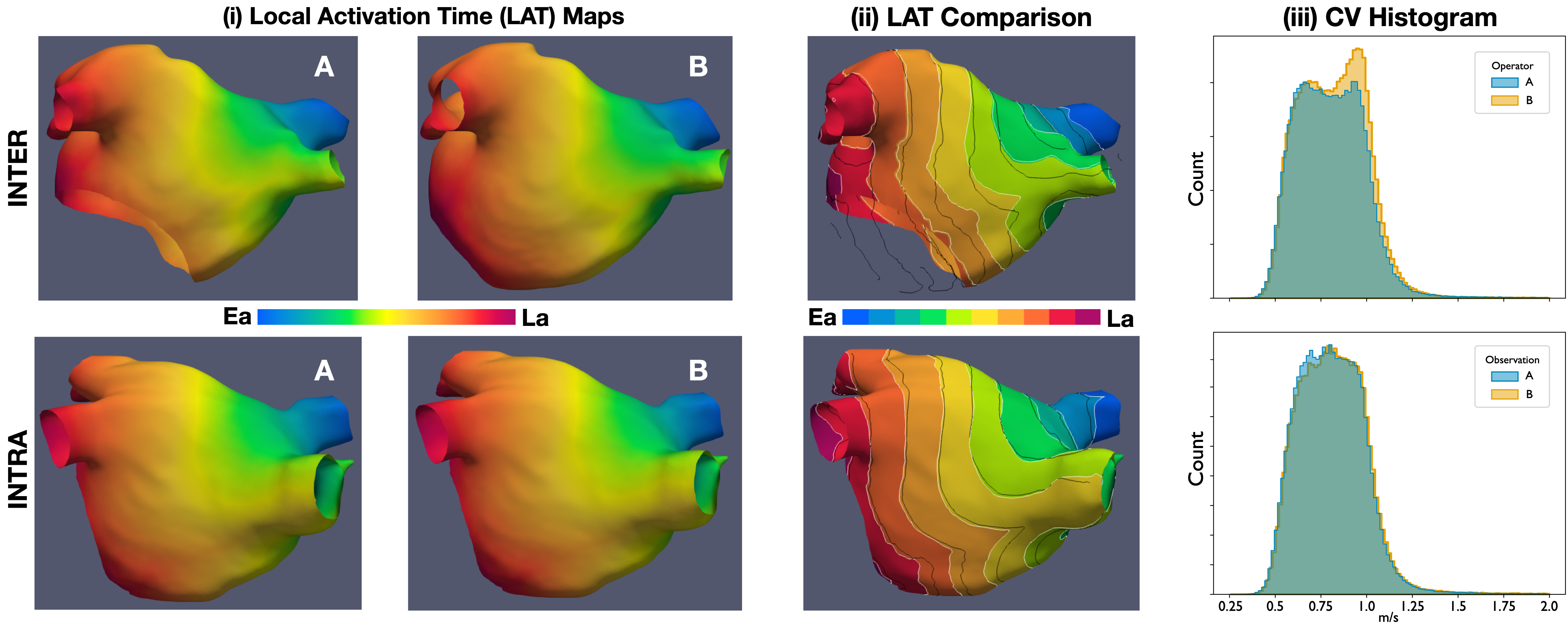}
  \caption{
    Example of simulated Local Activation Time (LAT) Maps and Conduction Velocity histograms. 
    \textbf{Top row}: comparison between operator A vs operator B for inter-operator variability.
    \textbf{Bottom row}: comparison between observation A vs observation B from one of the operators.
    \textbf{Columns}. 
    (i) Local Activation Time Maps shows two observations of the same case, 
    inter or intra depending on the row.
    (ii) LAT Comparison shows the LAT map from A with white contours, 
    the contours of B's LAT are superimposed in black. 
    (iii) The distribution of Conduction Velocity (CV) is shown for both A and B. 
    Ea, Early activation; La, Late activation; CV, Conduction Velocity; 
    LAT, Local Activation Time.
    }
    \label{fig:res-sims}
\end{figure}

Figure~\ref{fig:res-sims} shows two examples of simulation outputs,
corresponding for a case of inter and intra-operator variability. 
Each LAT map is presented with a colourmap ranging from early activation (Ea) to late activation (La).
A qualitative comparison between local activation time (LAT) maps is shown in column (iii),
where the contours are shown overlapped to appreciate visually the differences in propagation patterns. 
The histograms corresponding to each observation local conduction velocity are shown overlapped.
In the case for inter-operator variability, meshes have a notably different geometry around the mitral valve.
This can be appreciated more closely in column (iii) of Figure\ref{fig:res-sims}.

\section{Discussion}\label{sec:discussion}

This work describes an extension of CemrgApp to create simulation-ready
meshes from a pair of CMR scans.
CemrgApp is a software platform designed to be extended through standalone plugins.
We presented a model reproducibility study, where 50 cases were distributed amongst
five users to generate 100 models of the left atrium with two sets of fibre orientations.
Before analysing the reproducibility results, we discuss two general points.
First, the processing of cases between semi-automatic and manual cases was 51 to 49, respectively.
It is worth noting that the high number of cases where users decided to use the manual workflow
was mainly due to labelling errors rather than an incorrect segmentation of the atrial body.
In spite of this, the time to process a case was substantially reduced from
4.5 hours (15GB, 4 cores) in the work by~\cite{Roney2020-cinc},
to the median value of 16 minutes and 43 seconds in this study. 
Even considering only the use cases where a laptop was used, the time to process 
a case was reduced to a median value of 17 minutes and 32 seconds. 
It does not appear the hardware constituted a bottleneck in the processing time.
Second, the quality control stage carried out after collecting the data from users,
where only two instances of user error were found and corrected.
It is important to note that the quality control stage described in Section
\ref{sec:results} was carried out without introducing bias in the analysis.
As discussed in the supplementary material,
(i) fixing the labels assigned to the pulmonary veins is an automatic process, 
which has no impact on the user's decision on the position or shape of the veins; 
(ii) clipping the pulmonary veins in the scar projection mesh utilises the 
user's pre-defined clippers.


\paragraph{Shape Agreement.}
Mean and median distances were within 1-2mm,
which is close to the image resolution for all atrial structures.
Left pulmonary veins presented minimum distance values higher than 1mm in some cases.
The main reason was the variability when deciding where the vein starts and where to
clip the vein.
For the \emph{worst-case-scenario} calculation of the Hausdorff distance,
the larger problems were in the left atrial body and appendage, where
differences were larger than 5mm.
The mean Hausdorff distance of 9.5mm was comparable to other segmentation studies,
reporting mean Hausdorff distances of 20mm and 4.2mm
from~\cite{Li2020_hd} and~\cite{Ghosh2020_hd}, respectively.
It is worth noting that the studies were of fully automatic CNN-based methods.
Furthermore,~\cite{Li2020_hd} reported a Hausdorff distance of 36.4mm from a
competing U-Net-based segmentation, whilst the maximum Hausdorff distance
in this study was of 29.9mm.
The main problem with the left atrial body was the clipping of a mitral valve,
which varied substantially.
The largest differences were found in the inter-operator variability comparisons.

\paragraph{Fibrosis Agreement.}
Compared to the reproducibility measurements by~\cite{Sim2019-reproducibility},
our fibrosis agreement was overall higher at an ICC of 0.909 for inter and
an ICC of 0.999 for intra-operator variability.
If the thresholds are analysed independently, then the results become comparable
to our previous study, with the lowest ICC in the inter-operator variability at
an IIR of 1.32.
From Figure~\ref{fig:res-fibrosis-agreement} (left), two data points stand out,
corresponding to measurements at IIR=1.2 and IIR=1.32.
For context,~\cite{Sim2019-reproducibility}
investigated reproducibility of manually segmenting the
atrial body, identifying and clipping of the mitral valve,
pulmonary veins and left atrial appendage;
intra-class correlation (ICC) coefficients of 0.88 for inter
and 0.94 for intra-operator variability were reported.

\paragraph{Fibre Orientation Agreement.}
In both inter and intra-operator variability tests,
the distribution of fibre orientations appear in good agreement.
In the distributions presented on Figure~\ref{fig:res-fibre-agreement},
where 1 corresponds to perfect agreement, the percentages of angle
between $\pm 22.5^{\circ}$ are 60.63\% in inter-operator variability cases,
and 71.77\% in the intra-operator variability cases.
Put into context, in~\cite{Roney2021-atlas} reported fibre
agreement of approximately 33.36\% ($\pm 6.88$),
when comparing between the fibre fields 1 and the
rule-based Labarthe field, used also in this work.

\paragraph{Simulation Results.}
Our results support that the distribution of
total activation time is similar for both inter and intra comparisons.
First, the Wilcoxon rank test for total activation time was not statistically significant,
indicating that the assumption of equal means could not be rejected for
either inter or intra comparisons.
Furthermore, the medians and inter-quartile range were very similar
for both inter and intra-observer variations.
This is comparable to the differences between different atrial fibre fields,
reported by~\cite{Roney2021-atlas},
which would make inter and intra-observer uncertainty on the scale of inherent
uncertainty due to the inability to measure the atrial fibres.
A similar case occurred with the mean conduction velocity, 
although the individual comparisons were significantly different based on the output of 
each individual t-test.
The examples in Figure \ref{fig:res-sims} show a closer resemblance of the contour lines
between early activation (Ea) and late activation (La) in the intra-operator variability outputs.
Compared to the inter-operator variability, which shows a notably different shape.
This result is consistent with the other agreement metrics shown in this work,
such as shape, fibrosis, or fibre agreements.
The comparison metrics presented show a low correlation between the
PS maps, and only a \emph{modest} similarity index.
We note that with increased complexity, the risk of adding in variation increases.
Thus, even when local activation times will generate consistent results,
fibrillation results are invariably more prone to variation
as it constitutes a more complex simulation.
In~\cite{Roney2021-atlas}, the mean correlation of PS maps between
fibre fields ranged low from 0.14-0.44, and results varied based on
fibre field and anatomy.
It is worth noting that we performed only a single simulation of AF,
however aggregating PS maps across multiple pacing protocols~\cite{Boyle2021_af,Loewe2021_af}
may lead to more consistent results.
Longer simulation times may also lead to more
stable results~\cite{Corrado2021_ps}.

\subsection[Limitations]{Limitations}\label{sec:discussion-limitations}

\paragraph{Limitations of the CemrgApp Pipeline.}
The largest difference in shape came from the positions of the mitral valve,
when selected manually. This could be overcome by enhancing the manual variant
to keep the mitral valve segmentation and use it, removing user input from it.
At the moment this is only possible in the automatic variant of the application.
This limitation impacts the generation of the universal atrial coordinates,
since they depend on the geometry of the meshes, which in turn affects the fibre mapping and simulation outputs.
We are currently developing an extension to the universal atrial coordinates that
removes the requirement to label the mesh.

\paragraph{Limitations to the Universal Atrial Coordinates Implementation.}
The Universal Atrial Coordinates pipeline assumes there are four pulmonary veins.
For the CemrgApp plugin, we incorporated tools to ignore smaller veins detected.
A possible extension could be to mark the location of these extra veins in the
universal atrial coordinates,
to allow a more detailed investigation in the impact these structures.

\section[Conclusion]{Conclusion}
\label{sec:conclusion}
We have presented\footnote{This preprint has been submitted to the Journal Computers in Biology and Medicine.}
 an open-source, pipeline to produce models of the left atrium starting
from a pair of CMR scan(s) through to a simulation-ready mesh with
(1) estimated fibrosis, and (2) fibre orientations projected onto the surface of the mesh.
We produced 100 models that test inter and intra-operator variability of the pipeline (split 60/40).
Although there were notable differences, starting in the shape agreement metrics, 
which propagated errors down the pipeline, both inter and intra-operator variability
was comparable with uncertainty in atrial models due to image resolution
or the use of estimated fibre orientations.

\paragraph{Practical Implications.}
(i) Patient-specific computational models of the heart are increasingly been used 
to develop and guide clinical therapies~\cite{pras2018-intro-regulatory}.
The software pipeline we have developed aims to support upcoming frameworks 
for the generation of patient-specific atrial models, which have an impact on the 
development of personalised therapies for atrial fibrillation and other illnesses.
(ii) The uncertainties presented in model anatomy, fibres and simulations provide a 
context for interpreting of simulation study results.
Thus, this work offers the first reproducibility study as a potential initial template for reporting 
simulation-study reproducibility, thus providing a benchmark for future improvements in 
model creation.
(iii) The software has a low barrier to entry and a low learning curve, 
making it accessible to a wide range of users.
Users adapted to the software pipeline quickly with minimal training consisting of up to an hour 
session and resources like instructional videos and a standard operating procedure document, 
available as supplementary material.
The software is fully open-source, can be run in a standard laptop computer in a shorter time, 
and its methodologies are standardised. All of these reasons are important considerations for 
clinical applications.

\paragraph{Final Remarks.}
Overall, we consider this software pipeline to represent a subtantial contribution to the 
development of patient-specific computational models of the heart, 
which will facilitate the transition towards the adoption of computational models into 
clinical applications and pave the way for more research, with larger cohorts. 

\section*{Access to Code, Binaries, and Documentation}
The version of CemrgApp with the plugin developed for this work is hosted on Github under
commit number \texttt{0539e31}, which at the time of writing
can be accessed at \url{https://github.com/CemrgAppDevelopers/CemrgApp/tree/0539e31}.
Binaries for Windows, Linux (Ubuntu) and macOS (intel) can be made available upon request.
Besides the standard operating procedure document submitted as supplementary material, 
tutorial videos have been uploaded to Youtube for the automatic pipeline and manual pipelines 
at the respective urls
\url{https://youtu.be/zU_czEPaCIs}, and
\url{https://youtu.be/G4G4y-QuVV4}.
\bibliography{myrefs} 

\begin{thebibliography}{10}

\bibitem{pras2018-intro-regulatory}
Tina~M. Morrison, Pras Pathmanathan, Mariam Adwan, and Edward Margerrison.
\newblock Advancing regulatory science with computational modeling for medical
  devices at the fda's office of science and engineering laboratories.
\newblock {\em Frontiers in Medicine}, 5, 2018.

\bibitem{pras2021-intro-credibility}
Bahram Parvinian, Ramin Bighamian, Christopher~George Scully, Jin-Oh Hahn, and
  Pras Pathmanathan.
\newblock Credibility assessment of a subject-specific mathematical model of
  blood volume kinetics for prediction of physiological response to hemorrhagic
  shock and fluid resuscitation.
\newblock {\em Frontiers in Physiology}, 12, 2021.

\bibitem{Niederer2011_electrophysiology}
Steven~A. Niederer, Eric Kerfoot, Alan~P. Benson, Miguel~O. Bernabeu, Olivier
  Bernus, Chris Bradley, Elizabeth~M. Cherry, Richard Clayton, Flavio~H.
  Fenton, Alan Garny, Elvio Heidenreich, Sander Land, Mary Maleckar, Pras
  Pathmanathan, Gernot Plank, José~F. Rodríguez, Ishani Roy, Frank~B. Sachse,
  Gunnar Seemann, Ola Skavhaug, and Nic~P. Smith.
\newblock Verification of cardiac tissue electrophysiology simulators using an
  <i>n</i>-version benchmark.
\newblock {\em Philosophical Transactions of the Royal Society A: Mathematical,
  Physical and Engineering Sciences}, 369(1954):4331--4351, 2011.

\bibitem{pras2014_verification}
Pras Pathmanathan and Richard~A. Gray.
\newblock Verification of computational models of cardiac electro-physiology.
\newblock {\em International Journal for Numerical Methods in Biomedical
  Engineering}, 30(5):525--544, 2014.

\bibitem{sander2015}
Sander Land, Viatcheslav Gurev, Sander Arens, Christoph~M. Augustin, Lukas
  Baron, Robert Blake, Chris Bradley, Sebastian Castro, Andrew Crozier, Marco
  Favino, Thomas~E. Fastl, Thomas Fritz, Hao Gao, Alessio Gizzi, Boyce~E.
  Griffith, Daniel~E. Hurtado, Rolf Krause, Xiaoyu Luo, Martyn~P. Nash, Simone
  Pezzuto, Gernot Plank, Simone Rossi, Daniel Ruprecht, Gunnar Seemann,
  Nicolas~P. Smith, Joakim Sundnes, J.~Jeremy Rice, Natalia Trayanova, Dafang
  Wang, Zhinuo Jenny~Wang, and Steven~A. Niederer.
\newblock Verification of cardiac mechanics software: benchmark problems and
  solutions for testing active and passive material behaviour.
\newblock {\em Proceedings of the Royal Society A: Mathematical, Physical and
  Engineering Sciences}, 471(2184):20150641, 2015.

\bibitem{Pathmanathan2018}
Pras Pathmanathan and Richard~A. Gray.
\newblock Validation and trustworthiness of multiscale models of cardiac
  electrophysiology.
\newblock {\em Frontiers in Physiology}, 9, 2018.

\bibitem{moon2002-intro-reprod}
Frank Grothues, Gillian~C Smith, James~C.C Moon, Nicholas~G Bellenger, Peter
  Collins, Helmut~U Klein, and Dudley~J Pennell.
\newblock Comparison of interstudy reproducibility of cardiovascular magnetic
  resonance with two-dimensional echocardiography in normal subjects and in
  patients with heart failure or left ventricular hypertrophy.
\newblock {\em The American Journal of Cardiology}, 90(1):29--34, 2002.

\bibitem{alabed2022-intro-mri}
Samer Alabed, Ahmed Maiter, Mahan Salehi, Aqeeb Mahmood, Sonali Daniel, Sam
  Jenkins, Marcus Goodlad, Michael Sharkey, Michail Mamalakis, Vera Rakocevic,
  Krit Dwivedi, Hosamadin Assadi, Jim~M. Wild, Haiping Lu, Declan~P. O’Regan,
  Rob~J. van~der Geest, Pankaj Garg, and Andrew~J. Swift.
\newblock Quality of reporting in ai cardiac mri segmentation studies – a
  systematic review and recommendations for future studies.
\newblock {\em Frontiers in Cardiovascular Medicine}, 9, 2022.

\bibitem{Lee2019_gender}
{Angela Wing Chung} Lee, {Declan P.} O'Regan, {Justin Simon} Gould, {Baldeep
  Singh} Sidhu, Benjamin Sieniewicz, Gernot Plank, {David R} Warriner, Pablo
  Lamata, {Christopher Aldo} Rinaldi, and {Steven Alexander} Niederer.
\newblock Sex-dependent qrs guidelines for cardiac resynchronization therapy
  using computer model predictions.
\newblock {\em Biophysical Journal}, 117(12):2375--2381, December 2019.

\bibitem{Roney2022-predicting}
{Caroline H.} Roney, Iain Sim, Jin Yu, Marianne Beach, Arihant Mehta, {Jose
  Alonso} Solis-Lemus, Irum Kotadia, John Whitaker, Cesare Corrado, Orod
  Razeghi, Edward Vigmond, {Sanjiv M} Narayan, Mark O'Neill, {Steven E.}
  Williams, and {Steven A.} Niederer.
\newblock Predicting atrial fibrillation recurrence by combining population
  data and virtual cohorts of patient-specific left atrial models.
\newblock {\em Circulation. Arrhythmia and electrophysiology}, January 2022.

\bibitem{bifulco2021_90models}
Savannah~F Bifulco, Griffin~D Scott, Sakher Sarairah, Zeinab Birjandian,
  Caroline~H Roney, Steven~A Niederer, Christian Mahnkopf, Peter Kuhnlein,
  Marcel Mitlacher, David Tirschwell, WT~Longstreth, Nazem Akoum, and Patrick~M
  Boyle.
\newblock Computational modeling identifies embolic stroke of undetermined
  source patients with potential arrhythmic substrate.
\newblock {\em eLife}, 10:e64213, May 2021.
\newblock Publisher: eLife Sciences Publications, Ltd.

\bibitem{Razeghi2020-cemrgapp}
Orod Razeghi, Jos\'e~Alonso Sol\'is-Lemus, Angela W.~C. Lee, Rashed Karim,
  Cesare Corrado, Caroline~H. Roney, Adelaide de~Vecchi, and Steven~A.
  Niederer.
\newblock {CemrgApp}: {An} interactive medical imaging application with image
  processing, computer vision, and machine learning toolkits for cardiovascular
  research.
\newblock {\em SoftwareX}, 12:100570, July 2020.

\bibitem{Sim2019-reproducibility}
Iain Sim, Orod Razeghi, Rashed Karim, Henry Chubb, John Whitaker, Louisa
  O'Neill, Rahul~K. Mukherjee, Caroline~H. Roney, Reza Razavi, Matthew Wright,
  Mark O'Neill, Steven Niederer, and Steven~E. Williams.
\newblock Reproducibility of atrial fibrosis assessment using {CMR} imaging and
  an open source platform.
\newblock {\em JACC: Cardiovascular Imaging}, 12(10):2076--2077, October 2019.

\bibitem{oneill2019-pulmonary}
Louisa O'Neill, Rashed Karim, Rahul~K. Mukherjee, John Whitaker, Iain Sim,
  James Harrison, Orod Razeghi, Steven Niederer, Tevfik Ismail, Matthew Wright,
  Mark~D. O'Neill, and Steven~E. Williams.
\newblock Pulmonary vein encirclement using an {Ablation} {Index}-guided
  point-by-point workflow: cardiovascular magnetic resonance assessment of left
  atrial scar formation.
\newblock {\em EP Europace}, 21(12):1817--1823, December 2019.

\bibitem{chubb2018-reproducibility}
Henry Chubb, Rashed Karim, S\'ebastien Roujol, Marta Nu\~{n}ez Garcia,
  Steven~E. Williams, John Whitaker, James Harrison, Constantine Butakoff,
  Oscar Camara, Amedeo Chiribiri, Tobias Schaeffter, Matthew Wright, Mark
  O'Neill, and Reza Razavi.
\newblock The reproducibility of late gadolinium enhancement cardiovascular
  magnetic resonance imaging of post-ablation atrial scar: a cross-over study.
\newblock {\em Journal of Cardiovascular Magnetic Resonance}, 20(1):21, March
  2018.

\bibitem{hopman2021-quantification}
Luuk H G~A Hopman, Pranav Bhagirath, Mark~J Mulder, Iris~N Eggink, Albert~C van
  Rossum, Cornelis~P Allaart, and Marco J~W Götte.
\newblock Quantification of left atrial fibrosis by {3D} late
  gadolinium-enhanced cardiac magnetic resonance imaging in patients with
  atrial fibrillation: impact of different analysis methods.
\newblock {\em European Heart Journal - Cardiovascular Imaging},
  23(9):1182--1190, November 2021.

\bibitem{hopman2022_ra}
Luuk H G~A Hopman, Julia~E Visch, Pranav Bhagirath, Anja~M van~der Laan, Mark~J
  Mulder, Orod Razeghi, Michiel J~B Kemme, Steven~A Niederer, Cornelis~P
  Allaart, and Marco J~W Götte.
\newblock Right atrial function and fibrosis in relation to successful atrial
  fibrillation ablation.
\newblock {\em European Heart Journal - Cardiovascular Imaging}, page jeac152,
  August 2022.

\bibitem{Razeghi2020-cnn}
Orod Razeghi, Iain Sim, Caroline~H. Roney, Rashed Karim, Henry Chubb, John
  Whitaker, Louisa O'Neill, Rahul Mukherjee, Matthew Wright, Mark O'Neill,
  Steven~E. Williams, and Steven Niederer.
\newblock Fully automatic atrial fibrosis assessment using a multilabel
  convolutional neural network.
\newblock {\em Circulation. Cardiovascular Imaging}, 13(12):e011512, December
  2020.

\bibitem{Roney2019-uac}
Caroline~H. Roney, Ali Pashaei, Marianna Meo, R\'emi Dubois, Patrick~M. Boyle,
  Natalia~A. Trayanova, Hubert Cochet, Steven~A. Niederer, and Edward~J.
  Vigmond.
\newblock Universal atrial coordinates applied to visualisation, registration
  and construction of patient specific meshes.
\newblock {\em Medical Image Analysis}, 55:65--75, July 2019.

\bibitem{Roney2021-atlas}
Caroline~H. Roney, Rokas Bendikas, Farhad Pashakhanloo, Cesare Corrado,
  Edward~J. Vigmond, Elliot~R. McVeigh, Natalia~A. Trayanova, and Steven~A.
  Niederer.
\newblock Constructing a {Human} {Atrial} {Fibre} {Atlas}.
\newblock {\em Annals of Biomedical Engineering}, 49(1):233--250, January 2021.

\bibitem{Roney2020-cinc}
{Caroline H.} Roney, Marianne Beach, Arihant Mehta, Iain Sim, Cesare Corrado,
  Rokas Bendikas, {Jose A.} Solis-Lemus, Orod Razeghi, John Whitaker, {Louisa
  O.} O'Neill, Gernot Plank, Edward Vigmond, {Steven E.} Williams, {Mark D.}
  O'Neill, and {Steven A.} Niederer.
\newblock {\em Constructing Virtual Patient Cohorts for Simulating Atrial
  Fibrillation Ablation}, pages 1--4.
\newblock Computing in Cardiology. IEEE Computer Society, United States,
  September 2020.

\bibitem{Boyle2021_af}
Patrick~M Boyle, Alexander~R Ochs, Rheeda~L Ali, Nikhil Paliwal, and Natalia~A
  Trayanova.
\newblock {Characterizing the arrhythmogenic substrate in personalized models
  of atrial fibrillation: sensitivity to mesh resolution and pacing protocol in
  AF models}.
\newblock {\em EP Europace}, 23(Supplement 1):i3--i11, 03 2021.

\bibitem{Boyle2019_optima}
Patrick~M. Boyle, Tarek Zghaib, Sohail Zahid, Rheeda~L. Ali, Dongdong Deng,
  William~H. Franceschi, Joe~B. Hakim, Michael~J. Murphy, Adityo Prakosa,
  Stefan~L. Zimmerman, Hiroshi Ashikaga, Joseph~E. Marine, Aravindan
  Kolandaivelu, Saman Nazarian, David~D. Spragg, Hugh Calkins, and Natalia~A.
  Trayanova.
\newblock Computationally guided personalized targeted ablation of persistent
  atrial fibrillation.
\newblock {\em Nature Biomedical Engineering}, 3:870--879, 2019.

\bibitem{Plank2021-opencarp}
Gernot Plank, Axel Loewe, Aurel Neic, Christoph Augustin, Yung-Lin Huang,
  Matthias A.~F. Gsell, Elias Karabelas, Mark Nothstein, Anton~J. Prassl, Jorge
  S\'anchez, Gunnar Seemann, and Edward~J. Vigmond.
\newblock The {openCARP} simulation environment for cardiac electrophysiology.
\newblock {\em Computer Methods and Programs in Biomedicine}, 208:106223,
  September 2021.

\bibitem{Neic2020-meshtool}
Aurel Neic, Matthias~A.F. Gsell, Elias Karabelas, Anton~J. Prassl, and Gernot
  Plank.
\newblock Automating image-based mesh generation and manipulation tasks in
  cardiac modeling workflows using meshtool.
\newblock {\em SoftwareX}, 11:100454, 2020.

\bibitem{Khurram2014_iir}
Irfan~M. Khurram, Roy Beinart, Vadim Zipunnikov, Jane Dewire, Hirad
  Yarmohammadi, Takeshi Sasaki, David~D. Spragg, Joseph~E. Marine, Ronald~D.
  Berger, Henry~R. Halperin, Hugh Calkins, Stefan~L. Zimmerman, and Saman
  Nazarian.
\newblock Magnetic resonance image intensity ratio, a normalized measure to
  enable interpatient comparability of left atrial fibrosis.
\newblock {\em Heart Rhythm}, 11(1):85--92, 2014.

\bibitem{Benito2016_iir}
Eva~M. Benito, Alicia Carlosena-Remirez, Eduard Guasch, Susana Prat-González,
  Rosario~J. Perea, Rosa Figueras, Roger Borràs, David Andreu, Elena Arbelo,
  J.~Maria Tolosana, Felipe Bisbal, Josep Brugada, Antonio Berruezo, and Lluis
  Mont.
\newblock {Left atrial fibrosis quantification by late gadolinium-enhanced
  magnetic resonance: a new method to standardize the thresholds for
  reproducibility}.
\newblock {\em EP Europace}, 19(8):1272--1279, 12 2016.

\bibitem{merkel2014docker}
 Dirk Merkel.
\newblock
  Docker: lightweight linux containers for consistent development and deployment.
\newblock {\em Linux journal}, 2014(239):2, 2014.

\bibitem{labarthe2016_model}
Simon Labarthe, Jason Bayer, Yves Coudière, Jacques Henry, Hubert Cochet,
  Pierre Jaïs, and Edward Vigmond.
\newblock {A bilayer model of human atria: mathematical background,
  construction, and assessment}.
\newblock {\em EP Europace}, 16(suppl 4):iv21--iv29, 11 2014.

\bibitem{Acosta-Hausdorff}
F.~Commandeur, J.~Velut, and O~Acosta.
\newblock A vtk algorithm for the computation of the hausdorff distance.
\newblock {\em The VTK Journal}, 07 2011.

\bibitem{Shrout-ICC}
P.~E. Shrout and J.~L. Fleiss.
\newblock Intraclass correlations: uses in assessing rater reliability.
\newblock {\em Psychological Bulletin}, 86(2):420--428, March 1979.

\bibitem{Elfving-ICC}
David Liljequist, Britt Elfving, and Kirsti Skavberg~Roaldsen.
\newblock Intraclass correlation – a discussion and demonstration of basic
  features.
\newblock {\em PLOS ONE}, 14:1--35, 07 2019.

\bibitem{courtemanche1998-ionic}
Marc Courtemanche, Rafael~J Ramirez, and Stanley Nattel.
\newblock Ionic mechanisms underlying human atrial action potential properties:
  insights from a mathematical model.
\newblock {\em American Journal of Physiology-Heart and Circulatory
  Physiology}, 275(1):H301--H321, 1998.

\bibitem{courtemanche1999ionic}
Marc Courtemanche, Rafael~J Ramirez, and Stanley Nattel.
\newblock Ionic targets for drug therapy and atrial fibrillation-induced
  electrical remodeling: insights from a mathematical model.
\newblock {\em Cardiovascular research}, 42(2):477--489, 1999.

\bibitem{roney2020silico}
Caroline~H Roney, Marianne~L Beach, Arihant~M Mehta, Iain Sim, Cesare Corrado,
  Rokas Bendikas, Jose~A Solis-Lemus, Orod Razeghi, John Whitaker, Louisa
  O’Neill, et~al.
\newblock In silico comparison of left atrial ablation techniques that target
  the anatomical, structural, and electrical substrates of atrial fibrillation.
\newblock {\em Frontiers in physiology}, 11:1145, 2020.

\bibitem{Lubrecht2021_latcv}
Jolijn~M Lubrecht, Thomas Grandits, Ali Gharaviri, Ulrich Schotten, Thomas
  Pock, Gernot Plank, Rolf Krause, Angelo Auricchio, Giulio Conte, and Simone
  Pezzuto.
\newblock {Automatic reconstruction of the left atrium activation from sparse
  intracardiac contact recordings by inverse estimate of fibre structure and
  anisotropic conduction in a patient-specific model}.
\newblock {\em EP Europace}, 23(Supplement 1):i63--i70, 03 2021.

\bibitem{Cantwell2015_cv}
C.D. Cantwell, C.H. Roney, F.S. Ng, J.H. Siggers, S.J. Sherwin, and N.S.
  Peters.
\newblock Techniques for automated local activation time annotation and
  conduction velocity estimation in cardiac mapping.
\newblock {\em Computers in Biology and Medicine}, 65:229--242, 2015.

\bibitem{Li2020_ps}
Xin Li, Tiago~P. Almeida, Nawshin Dastagir, María~S. Guillem, João Salinet,
  Gavin~S. Chu, Peter~J. Stafford, Fernando~S. Schlindwein, and G.~André Ng.
\newblock Standardizing single-frame phase singularity identification
  algorithms and parameters in phase mapping during human atrial fibrillation.
\newblock {\em Frontiers in Physiology}, 11, 2020.

\bibitem{solislemus2022_modrep_dataset}
Jose~Alonso Solis~Lemus, Tiffany Baptiste, Rosie Barrows, Charles Sillet, Ali
  Gharaviri, Giulia Raffaele, Orod Razeghi, Marina Strocchi, Iain Sim, Irum
  Kotadia, Neil Bodagh, Daniel O'Hare, Mark O'Neill, Steven Williams, Caroline
  Roney, and Steven Niederer.
\newblock Model reproducibility study on left atrial fibres, 2022.

\bibitem{Li2020_hd}
Lei Li, Xin Weng, Julia~A. Schnabel, and Xiahai Zhuang.
\newblock Joint left atrial segmentation and scar quantification based on a dnn
  with spatial encoding and shape attention.
\newblock In Anne~L. Martel, Purang Abolmaesumi, Danail Stoyanov, Diana Mateus,
  Maria~A. Zuluaga, S.~Kevin Zhou, Daniel Racoceanu, and Leo Joskowicz,
  editors, {\em Medical Image Computing and Computer Assisted Intervention --
  MICCAI 2020}, pages 118--127, Cham, 2020. Springer International Publishing.

\bibitem{Ghosh2020_hd}
Shrimanti Ghosh, Nilanjan Ray, Pierre Boulanger, Kumaradevan Punithakumar, and
  Michelle Noga.
\newblock Automated left atrial segmentation from magnetic resonance image
  sequences using deep convolutional neural network with autoencoder.
\newblock In {\em 2020 IEEE 17th International Symposium on Biomedical Imaging
  (ISBI)}, pages 1756--1760, 2020.

\bibitem{Loewe2021_af}
Luca Azzolin, Steffen Schuler, Olaf Dössel, and Axel Loewe.
\newblock A reproducible protocol to assess arrhythmia vulnerability in silico:
  Pacing at the end of the effective refractory period.
\newblock {\em Frontiers in Physiology}, 12, 2021.

\bibitem{Corrado2021_ps}
Cesare Corrado, Steven Williams, Caroline Roney, Gernot Plank, Mark O’Neill,
  and Steven Niederer.
\newblock {Using machine learning to identify local cellular properties that
  support re-entrant activation in patient-specific models of atrial
  fibrillation}.
\newblock {\em EP Europace}, 23(Supplement 1):i12--i20, 01 2021.

\end{thebibliography}
\bibliographystyle{unsrt} 



\addcontentsline{toc}{chapter}{Appendix}
\appendix

\section{Access to Code and Binaries}
The version of CemrgApp with the plugin developed for this work is hosted on Github under
commit number \texttt{0539e31}, which at the time of writing
can be accessed at \url{https://github.com/CemrgAppDevelopers/CemrgApp/tree/0539e31}.
Binaries for Windows, Linux (Ubuntu) and macOS (intel) can be made available upon request.

\section{Access to Data and Training Materials}
Models have been made available on Zenodo under the DOI \url{10.5281/zenodo.7433015}.
Besides the standard operating procedure document submitted as supplementary material, 
tutorial videos have been uploaded to Youtube for the automatic pipeline and manual pipelines 
at the respective urls
\url{https://youtu.be/zU_czEPaCIs}, and
\url{https://youtu.be/G4G4y-QuVV4}.

\section{Allocation and distribution of cases per user}
\begin{figure}[hbpt]
    \centering
    \includegraphics[width=\textwidth]{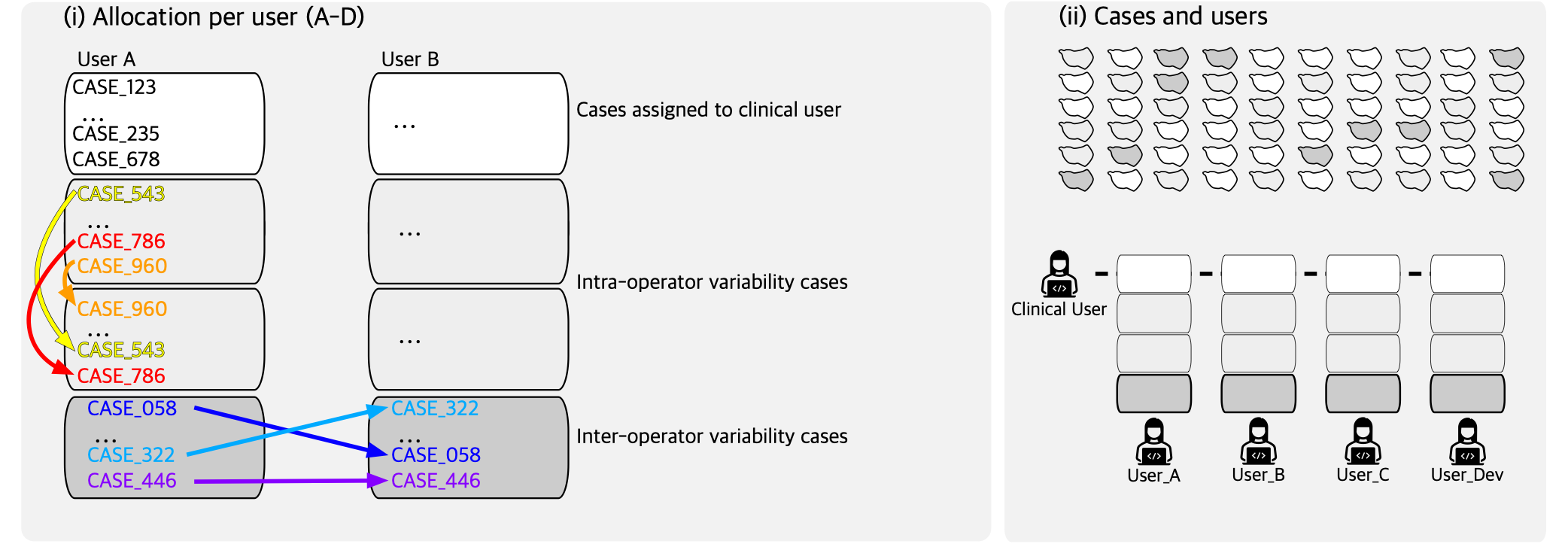}
    \caption[Allocation and distribution of cases per user.]
    {
      Allocation and distribution of cases per user and role of senior user.
    (a) Shows the allocation of cases per pair of users.
    Grayscale represents the category of case allocated.
    (b) Shows the distribution of case category from the pool of fifty.
    Twenty out of 50 were non-repeated cases (white),
    20 out of 50 were used for intra-operator variability, and
    10 out of 50 were used for inter-operator variability (dark grey).}
    \label{fig:mat-cases-per-user}
\end{figure}

Data were divided amongst five users: the developer (d), three novice users (abc),
and one senior user with more experience with medical images (s).
Only the developer had seen and screened each of the cases
before they were randomly assigned to each user.
The allocation and distribution of cases per user and role of senior user
are presented in Figure \ref{fig:mat-cases-per-user}, where an example of
data allocation is presented for a pair of users (from the abcd).
The role of the senior user (s) is shown on the right hand side of the Figure.
Each user (abcd) was assigned randomly 20 cases to process.
The data assigned to each user were categorised as follows:
(i) \textbf{10 cases} for \textbf{intra}-operator variability;
(ii) \textbf{5 cases} for \textbf{inter}-operator variability vs. another user (abcd);
and (iii) \textbf{5 cases} for inter-operator variability vs. senior user (s).

\section{Methods}
\begin{figure}[hbpt]
  \centering
  \includegraphics[width=\textwidth]{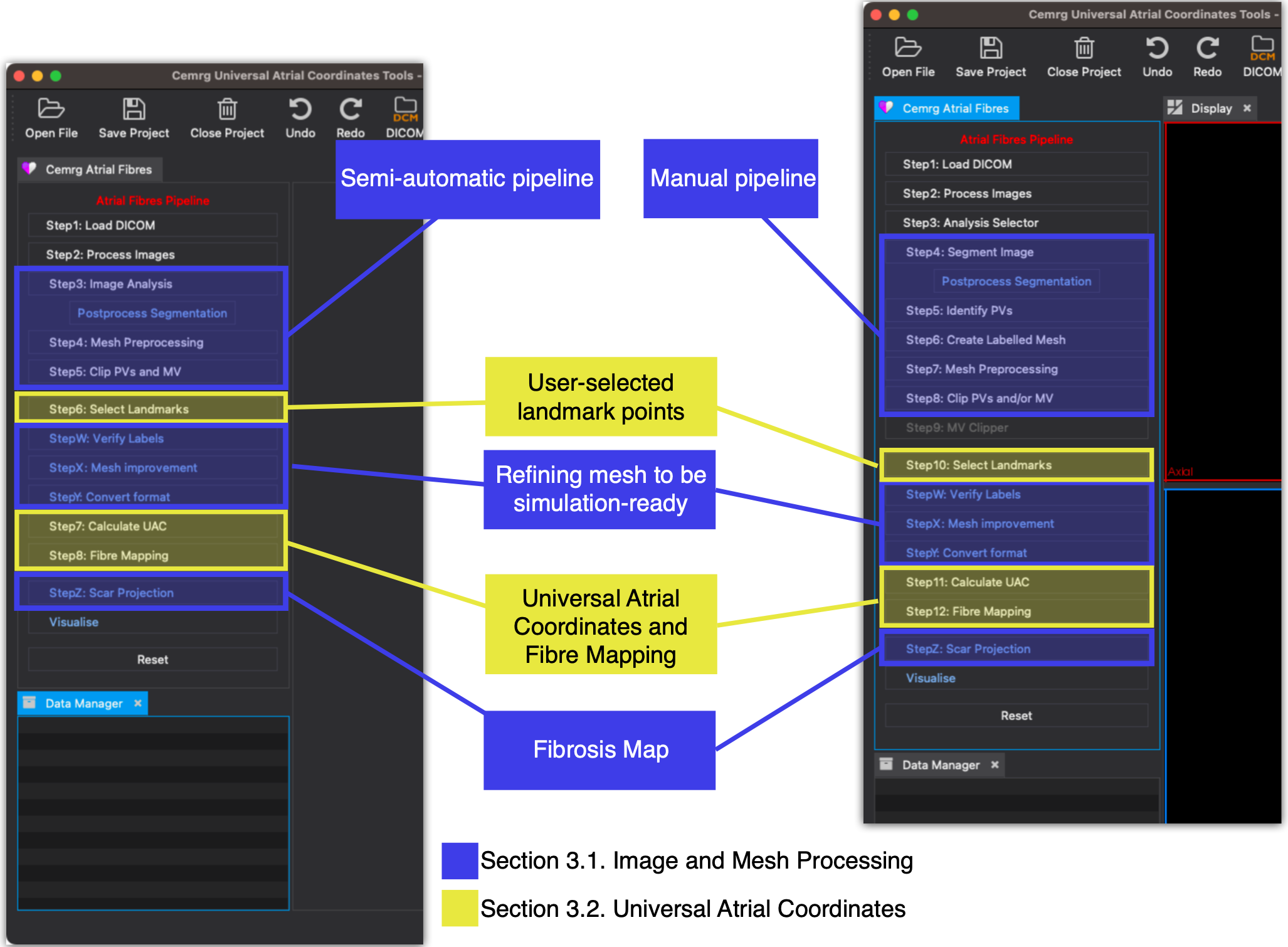}
  \caption{
    Representation of user interface (UI) elements in the CemrgApp plugin 
    and their connection to the description in the main text. 
    The UI elements are grouped into two categories: 
    (1) buttons to do Image and Mesh Processing (blue), and 
    (2) buttons to perform the Universal Atrial Coordinates (UAC) and 
    Fibre Mapping processing (yellow). 
    Two instances of the UI are shown as some of the buttons change depending 
    on the processing mode selected by the user. 
    Left: semi-automatic processing mode. Right: manual processing mode.
    Notice that the manual pipeline show more buttons, as the steps require 
    more user input. 
    Buttons are numbered automatically depending on the processing mode selected.
  }
  \label{fig:ui-elements}
\end{figure}

The CemrgApp plugin developed for this work is based on a series of buttons. 
Figure~\ref{fig:ui-elements} show a graphical representation of the buttons
and their conncetion to Sections 3.1 and 3.2 of the main text.
\subsection{Semi-automatic vs manual pipelines}
\begin{figure}[hbpt]
  \centering
  \includegraphics[width=\textwidth]{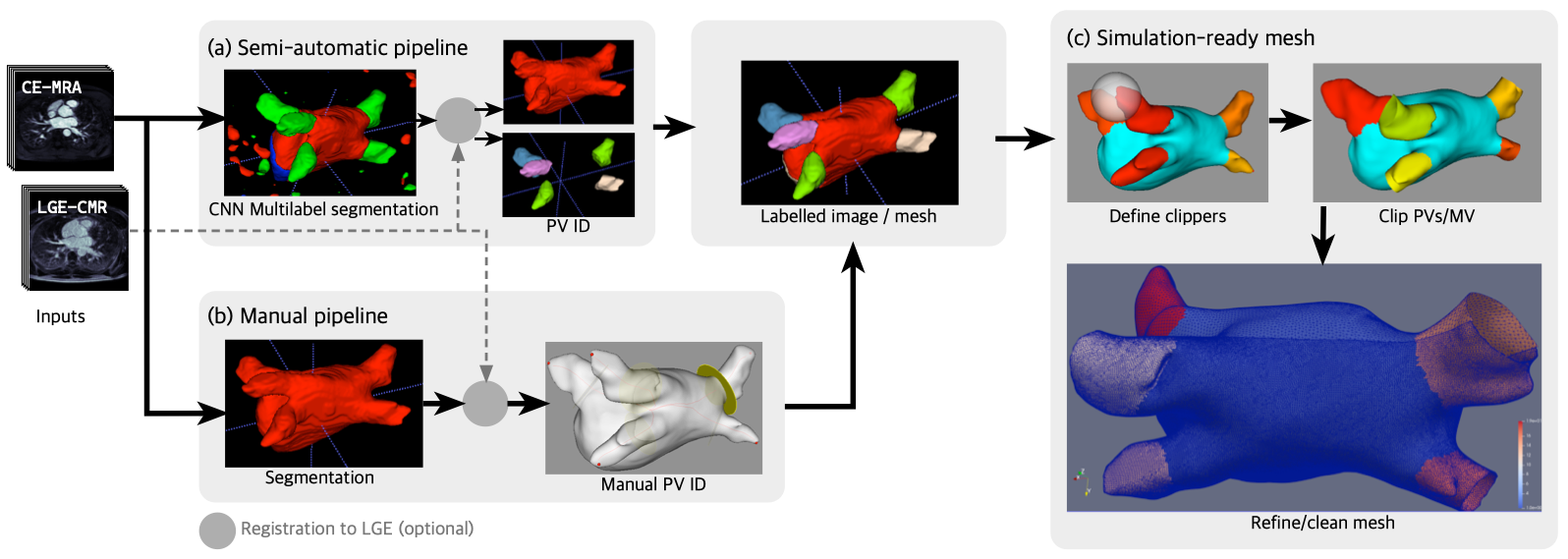}
  \caption{
    Distinction between semi-automatic and manual segmentation and pulmonary vein
    identification in CemrgApp module.
    From left to right, the input scans (MRA and LGE) are loaded into the system and segmented
    with a multi-label CNN-based automatic segmentation (a) or a manual segmentation (b).
    The automatic segmentation is from the work by~\cite{Razeghi2020-cnn}, it has three labels:
    atrial body, pulmonary veins, and mitral valve.
    The manual segmentation can be done by editing the single-label version of the
    CNN referenced before.
    The dotted arrow shows the moment in the pipeline where the segmentation --based on the MRA scan--
    is registered to the LGE scan.
    To create a simulation-ready mesh (c), the pulmonary veins are clipped by user-defined spheres at the
    distal end of the pulmonary veins. The mitral valve is also removed.
    Finally, the mesh is refined to an edge length of 0.3mm and rescaled to be in $\mu m$.
  }
  \label{fig:methods-im2surf}
\end{figure}

The CemrgApp plugin presented introduced two modes of operation to produce a labelled mesh:
\emph{semi-automatic} and \emph{manual}.
the semi-automatic pipeline is based on the work by~\cite{Razeghi2020-cnn}, 
which was trained on 207 cases achieving a dice score of $0.91 \pm 0.2$.
Figure \ref{fig:methods-im2surf} shows a graphical representation of the way both pipelines
integrate seamlessly into a mesh with labelled elements representing the different atrial
structures.
The diagram also shows the point where the MRA scan is registered with the LGE to
later perform the fibrosis identification.
Finally, the creation of a simulation-ready mesh is done by clipping the mesh at the distal
ends of the pulmonary veins and the mitral valve. Then, the mesh is refined to an edge length
of 0.3 $mm$ and scaled to be in $\mu m$.

\subsection{Mesh post-processing and refining}
\begin{figure}[hbpt]
    \centering
    \includegraphics[width=\textwidth]{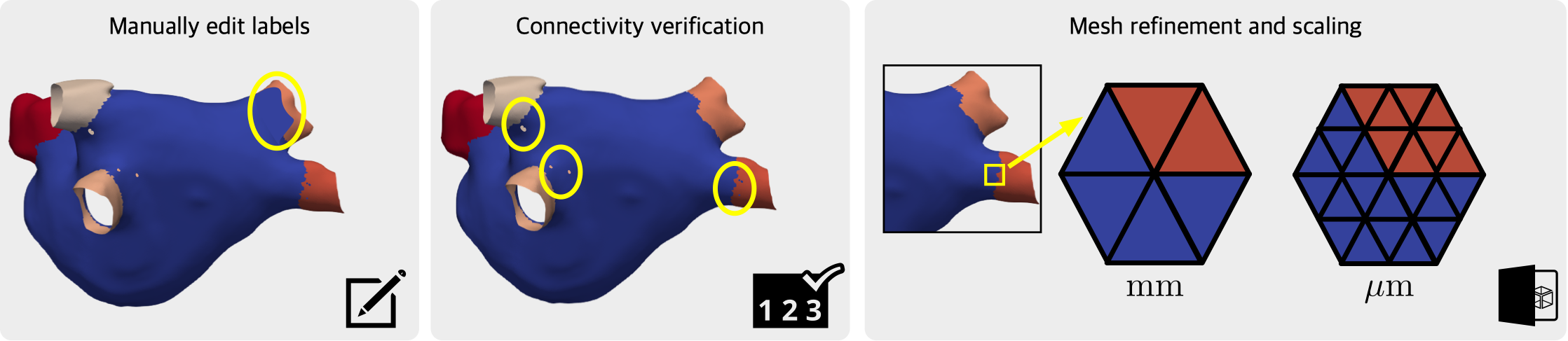}
    \caption[Mesh processing.]
    {
      Mesh post-processing and refining.
      Tools for manual editing (left) and automatic error-correction (centre)
      of labels in meshes are incorporated in the CemrgApp pipeline.
      Finally, mesh refining and cleaning (right) is done through meshtool, by~\cite{Neic2020-meshtool},
      which is incorporated in CemrgApp through openCARP's docker container.

    }
    \label{fig:methods-meshproc}
\end{figure}

Tools were developed as a part of the CemrgApp pipeline presented to manually edit labels and
to automatically correct errors in labelling, see Figure~\ref{fig:methods-meshproc}.
The users can manually edit labels by setting control points over the desired areas in the surface mesh,
then they select the label they wish to use to cover.
The user then presses a button,
which will define the shortest geodesic paths between the points, in the order they were selecter,
and assign the selected label onto all the points in the corridor and
their neighbouring elements in the mesh.
An automatic verification tool was also developed, which iterates through the labels in the mesh,
checking there are no \emph{islands} of elements with a different label than what they should be.
For example, in Figure~\ref{fig:methods-meshproc} (centre), mislabelled elements are highlighted.
These would be re-labelled automatically to the correct value.

\begin{figure}[hbpt]
    \centering
    \includegraphics[width=\textwidth]{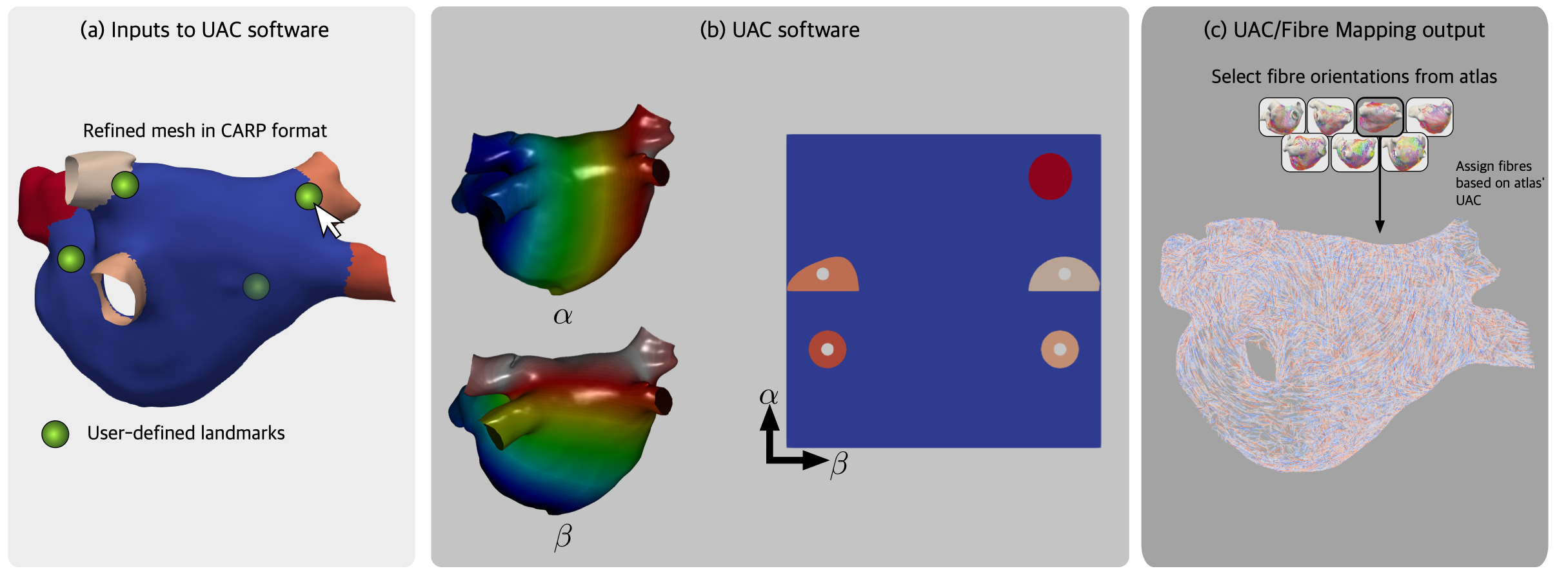}
    \caption{
      Universal Atrial Coordinates and Fibre Mapping Stages. 
      The diagram shows the steps from the user landmarks to the fibre mapping.
    }
    \label{fig:methods-uac}
\end{figure}

\subsection{Connection between CemrgApp and the Universal Atrial Coordinates docker container}
The user selects four landmark points, which guide the creation of
the Universal Atrial Coordinates (UAC)~\cite{Roney2019-uac}.
The UAC code has been packaged in a docker container
(\url{hub.docker.com/repository/docker/cemrg/uac}).
The universal atrial coordinates allow the mapping of fibre orientations from
an atlas~\cite{Roney2021-atlas}.
Detailed description of the UAC and fibre mapping algorithm are found in the
works~\cite{Roney2019-uac,Roney2021-atlas}.
Figure~\ref{fig:methods-uac} shows a diagram of the process from the user landmarks,
universal atrial coordinates, and fibre mapping on a particular case.

\section{Results}
\subsection{Quality control}
\begin{figure}[hbpt]
    \includegraphics[width=\textwidth]{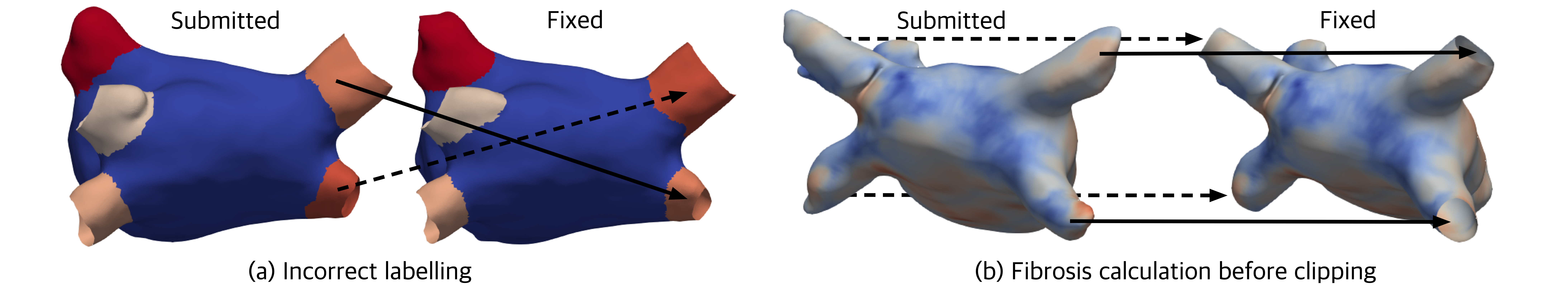}
    \caption[Cases identified during the Quality Control stage of this study.]
    {
    Cases identified during the Quality Control stage of this study.
    In both instances, the submitted and the fixed mesh are indicated
    with labels and arrows
    (a) The incorrect labelling shows the right superior and inferior
    pulmonary veins in incorrect order.
    In (b), the mesh has not been clipped at the moment of calculating
    the scar projection.
    }\label{fig:res-qc}
  \end{figure}

The 100 cases processed were screened for quality control.
Two cases presented a user problem which, if not addressed,
would have exclude them from the analysis.
Both cases were from the intra-observer variability sets.
The first, where the user identified the RSPV and RIPV in different order
during the mesh preprocessing stage of the \emph{semi-automatic pipeline}
described in Section 3.1.
The case was reprocessed by the same user from the starting pair of scans.
The second error was caused when another user calculated the Scar Projection
of the pipeline before clipping the mesh.
The surface mesh with the scar projection was clipped using the same clippers
saved from the corresponding stage in the pipeline.
The fix was performed by the developer, as the user's involvement was already
stored in the clipper positions and radii.
Figure~\ref{fig:res-qc} shows the two cases identified during the quality control
screening for reprocessing.

\subsection{Fibrosis Agreement}
Table~\ref{tab:fibrosis-icc} shows the detailed intra-class correlation (ICC) coefficient
results in detail for the fibrosis agreement reproducibility tests.
ICC was calculated specific to each IIR threshold calculated.
Confidence intervalsand p-values are reported.
In all cases, the p-value of the ICC is lower than the standard 0.05.

\begin{table}[h]
  \caption{Intra-class correlation (ICC) coefficient for
    inter and intra-observer tests in fibrosis score.
    ICC was calculated overall, and per each IIR threshold 0.97, 1.2, and 1.32.
    p-values and confidence intervals at 95\% are reported.
  }
  \label{tab:fibrosis-icc}
  \begin{tabular}{lrrrlrrr}
    \hline
    \textbf{Inter} & ICC & pval & CI95\% & \textbf{Intra} & ICC  & pval & CI95\% \\
    \hline
    Overall & 0.909 & 1.77e-9 & [0.81 0.96] & Overall & 0.999 & 2.23e-6 & [0.99 1.] \\
    0.97 & 0.987 & 3.56e-21 & [0.97 0.99] & 0.97 & 0.985 & 6.89e-4 & [0.85 1.] \\
    1.2 & 0.875 & 9.36e-8 & [0.74 0.94] & 1.2 & 0.999 & 3.03e-7 & [0.99 1.] \\
    1.32 & 0.851 & 7.06e-7 & [0.69 0.93] & 1.32 & 0.999 & 7.83e-7 & [0.99 1.] \\
    \hline
  \end{tabular}
\end{table}



\end{document}